\documentclass[referee,extra]{gji}
\usepackage{amsmath}
\usepackage{bm}
\usepackage[hidelinks]{hyperref}

\usepackage{color}
\definecolor{lightgray}{gray}{0.7}    
\definecolor{darkblue}{rgb}{0,0,0.7}
\definecolor{darkred}{rgb}{0.7,0,0}
\usepackage{soul}
\usepackage[normalem]{ulem}
\usepackage{graphicx}
\usepackage{lineno} 
\usepackage{amsmath,amssymb}
\usepackage{gensymb}
\usepackage[nodisplayskipstretch]{setspace}
\usepackage{caption}

\usepackage{titlesec}
\titleformat{\subsubsection}[runin]
{\normalfont\bfseries}{\thesubsubsection}{1em}{}
\begin{document}

\begin{center}
\textbf{\Large {Nucleation of laboratory earthquakes: quantitative analysis and scalings}} \\[20pt]
\textcolor{blue}{\small Manuscript submitted to J. Geophys. Res.}\\[20pt]

S. Marty$^{1}$, A. Schubnel$^{2}$, H. S. Bhat$^{2}$, J. Aubry$^{2}$, E. Fukuyama$^{3,4}$, S. Latour$^{5}$, S. Nielsen$^{6}$  and R. Madariaga$^{1}$

\begin{enumerate}
\small
\it
\itemsep0em
\item{Rock and Sediments mechanics laboratory, Penn State University, State College, USA}
\item{Laboratoire de G\'{e}ologie, \'{E}cole Normale Sup\'{e}rieure, CNRS-UMR 8538, PSL Research University, Paris, France.}
\item{National Research Institute for Earth Science and Disaster Resilience, Tsukuba, Ibaraki 305-0006, Japan}
\item{Department of Civil and Earth Resources Engineering, Kyoto University, Kyoto 615-8530, Japan}
\item{Université Toulouse III - Paul Sabatier, Toulouse, France}
\item{Department of Earth Sciences, Durham University, Durham, United Kingdom}
\end{enumerate}

\end{center}
\vspace{0.2cm}

\noindent{\bf 
Decades of seismological observations have highlighted the variability of foreshock occurrence prior to natural earthquakes, making thus difficult to track how earthquakes start. Here, we report on three stick-slip experiments performed on cylindrical samples of Indian metagabbro under upper crustal stress conditions (30-60 $MPa$). Acoustic emissions (AEs) were continuously recorded by 8 calibrated acoustic sensors during the experiments. Seismological parameters (moment magnitude, corner frequency and stress-drop) of the detected AEs  ($-8.8 \leq Mw \leq -7 $) follow the scaling law between moment magnitude and corner frequency that characterizes natural earthquakes. AE activity always increases towards failure and is found to be driven by along fault slip velocity. Consistently for all three experiments, the stacked AE foreshock sequences follow an inverse power-law of the time to failure (inverse Omori), with a characteristic Omori time $c$ inversely proportional to normal stress. AEs moment magnitudes increase towards failure, as manifested by a decrease in b-value from $\sim 1$ to $\sim 0.5$ at the end of the nucleation process. During nucleation, the averaged distance of foreshocks to mainshock also continuously decreases, highlighting the fast migration of foreshocks towards the mainshock epicenter location, and stabilizing at a distance from the latter compatible with the predicted Rate-and-State nucleation size. Importantly, we also show that the nucleation characteristic timescale scales inversely with applied normal stress and the expected nucleation size. Finally, the seismic component of the nucleation phase is orders of magnitude smaller than that of its aseismic component, which suggests that, in this experimental setting at least, foreshocks are the byproducts of a process almost fully aseismic. Nevertheless, the seismic/aseismic energy release ratio continuously increases during nucleation, highlighting that, if the nucleation process starts as a fully aseismic process, it evolves towards a cascading process at the onset of dynamic rupture.  

\begin{itemize}
{
\item Foreshock sequences follow an inverse Omori-law during which foreshock migration towards the final epicenter is observed.
\item The nucleation process scales inversely with normal stress and an experimental scaling is found for inverse Omori-law.  
\item The seismic/aseismic energy release continuously increases during nucleation, highlighting that the nucleation process starts as fully aseismic to evolve towards a cascading process. 
}
\end{itemize}
}

\section{Introduction}
The term "foreshocks" refers to small earthquakes that would occur nearby in time and space of a larger earthquake to come. \citet{papazachos1973time} made the observation that when a sufficient number of foreshock sequences were synchronized to the time of their respective mainshock and then stacked, the seismicity rate increases as an inverse power law of time when approaching the nucleation. This law, called "the inverse Omori law", had then provided a potential path to earthquake prediction. However, statistical models \citep{ogata1988statistical,helmstetter2003foreshocks} are able to reproduce most of the features attributed to foreshock sequences which was used as an argument to suggest that foreshocks merely reflect stochastic rather than physical processes. The outstanding question is thus whether or not earthquakes are preceded by a slow, emerging nucleation phase before propagating dynamically or start as small instabilities that may eventually grow bigger. These two opposite views are termed the "preslip" and the "cascade" models respectively \citep{ellsworth1995seismic,beroza1996properties}. In the latter scenario, the use of foreshocks as a predictive tool for the occurrence of a larger earthquake would be compromised.

At the scale of the Earth's lithosphere, numerous seismological observations have reported increasing foreshock activity preceding the occurrence of large earthquakes \citep{jones1976frequency,abercrombie1996occurrence,bouchon2011extended,kato2014multiple}. Foreshock activity preceding large subduction earthquakes has been found to correlate with the occurrence of slow slip transients in the region close to the hypocenter \citep{kato2012propagation,ruiz2014intense}. When examining the occurrence of foreshock sequences with respect to the geodynamic context, it has been demonstrated that faults subject to high-slip rates produce more foreshock sequences \citep{mcguire2005foreshock,bouchon2013long}. Moreover, compared with the ordinary seismicity, foreshocks present singular characteristics such as migration and acceleration prior to the mainshock \citep{marsan2014foreshock,kato2016accelerated}. Therefore, it has been argued that foreshocks are a by-product of the larger nucleation phase of the upcoming mainshock. Indeed, earthquakes are dynamic instabilities which result from the weakening of frictional properties of a seismogenic fault that has started to slip. The relation between on-fault friction and slip provides the theoretical frame to understand how earthquakes nucleate. Based on either slip weakening or rate-and-state friction laws, theoretical \citep{ida1972cohesive,campillo1997initiation,uenishi2003universal} and numerical models \citep{dieterich1992earthquake,rubin2005earthquake,ampuero2008earthquake} have demonstrated that before propagating dynamically, slip initially develops on a localized, slowly growing zone, which is defined as the nucleation zone. 

In the framework of rate-and-state friction laws, models that use laboratory derived friction parameters predict that earthquakes nucleate on short time and space scales, of the order of milliseconds and meters respectively \citep{lapusta2003nucleation,kaneko2008variability,fang2010effect}. This is a consequence of the characteristic slip distance $D_{c}$ (i.e. the length required for the friction to reach its residual value inferred from rock friction experiments being of the order of 1-100 $\mu m$). In the former case, detecting earthquakes nucleation from geodetic or seismological measurements would likely be unreachable. On the other hand, seismological observations have suggested that $D_{c}$ should be scale dependent \citep{ide1997determination,olsen1997three}, of the order of the centimeter at the scale of crustal earthquakes. The scaling of $D_{c}$ has been attributed to length scales inherent to the size of earthquakes such as long wavelength roughness of fault zones \citep{ohnaka2003constitutive} or gouge thickness \citep{marone1998laboratory}. If we consider that the critical slip distance involved during co-seismic slip is the same that governs earthquake nucleation \citep{cocco2009scaling}, this would imply nucleation processes to happen at much larger length and time scales. 

In the last forty years, a large number of rock fracture and/or friction experiments \citep{lockner1993role,ojala2004strain,schubnel2007fluid,thompson2009premonitory,mclaskey2014preslip,kwiatek2014seismic,passelegue2017influence} have revealed an important increase in the rate of acoustic emission (AE) triggering prior to failure and/or slip propagation , which has emphasized the possibility of earthquake forecasting at the laboratory scale \citep{johnson2021laboratory}. Dedicated stick-slip experiments have also supported the conceptual view of earthquake nucleation, whether it is for experiments conducted at low normal stress conditions on polymer materials \citep{latour2013characterization,nielsen2010experimental,selvadurai2015laboratory, guerin2019earthquake, Gvirtzman2021nucleation}, on crustal rocks under bi-axial (i.e. unconfined) conditions \citep{okubo1984effects, ohnaka1990characteristic,ohnaka2003constitutive,mclaskey2013foreshocks, fukuyama2018spatiotemporal} or triaxial (i.e. confined and hence higher normal stress) conditions \citep{passelegue2017influence, harbord2017earthquake,acosta2019can,aubry2020fault}. Experimental works have also investigated changes in the frequency-magnitude distribution (i.e. the $b$-value or slope of the Gutenberg-Richter earthquake frequency-magnitude power law statistical relationship) of AEs during stick-slip cycles \citep{main1989reinterpretation, sammonds1992role, lockner1993role, goebel2012identifying}. When the shear stress increases and the rupture is developing, a significant drop of the $b$-value has been reported, i.e. the ratio between large and small AEs increases \citep{goebel2013acoustic,riviere2018evolution,lei2018seismic}. This was thought to be driven by accelerating slip before dynamic rupture propagation. Consequently, this indicates that $b$-value changes could be used as a tool for seismic hazard assessment. 

Despite all these efforts made to understand the driving forces of foreshock occurrence, the physical processes that govern their triggering are still controversial. It thus appears necessary to further constrain the length and time scales over which earthquakes nucleate, as well as the possible relation between foreshock and pre-slip during the nucleation phase. Here we report on precursory AE sequences during stick-slip experiments conducted on metagabbro saw-cut samples and under crustal stress conditions (30, 45 and 60 MPa). The purpose of this study is to use generated precursory AEs as a proxy to investigate the dominant mechanisms that control foreshock dynamics. We purposefully concentrate on  stacked sequences of foreshocks, in order to highlight their general behavior, nevertheless keeping in mind the diversity and variability of the processes at play. Using calibrated acoustic sensors, AE seismological parameters (absolute moment magnitude, corner frequency, source size and stress drop) are estimated. AE features such as magnitude-frequency distribution, spatial distribution and temporal evolution towards failure are examined and interpreted. We find a scaling for the premonitory inverse Omori-law and finally, rely on absolute AE moment magnitudes to estimate the ratio between the seismic and the aseismic components of the pre-failure phase.

\section{Experimental procedure}
Experiments were conducted on room dry cylindrical samples of Indian metagabbro, 88 $mm$ long and 40 $mm$ diameter. Samples contained a saw-cut surface inclined at an angle $\theta$=$30^{\circ}$ with respect to the vertical axis (figure 1b.). Saw-cut surfaces were ground flat and then manually roughen with a \#240 sandpaper (average particle diameter 125 $\mu m$). The basic properties of metagabbro are as follows: P-wave, S-wave velocities and bulk density respectively equal to $c_p$ = 6.92 $km/s$, $c_s$ = 3.62 $km/s$ and  $\rho$ = 2980 $kg/m^{3}$ \citep{fukuyama2018spatiotemporal}. 

Saw-cut samples were loaded in a triaxial oil-medium loading cell (figure 1a). In each test: (i) the sample was placed in a 125 $mm$ long and 4 $mm$ thick viton jacket to isolate it from the confining fluid and positioned in the confining chamber, (ii) the confining pressure ($Pc$) was applied and (iii) the axial stress was increased by moving the vertical ram against the top of the sample at a constant speed (10 $\mu m/s$). Axial piston displacement and confining pressure were both independently servo-controlled. Axial stress and confining pressure were measured with external load cells with $10^{-3}$ MPa resolution and axial shortening was measured at the top of the axial piston using a LVDT displacement sensor with $\pm$ 0.1 $\mu m$ resolution. Confining pressure, axial stress and axial shortening were recorded at 10 Hz during the experiments. More details on the mechanical set-up can be found in \citet{schubnel2005damage}.

Samples were instrumented with an array of 8 acoustic sensors positioned on one side of the fault plane (figure 1b). Each acoustic sensor consisted of a cylindrical piezoelectric crystal (PZT - lead zirconate titanate) 5 mm in diameter and 0.5 mm thick (PI ceramic PI255) encapsulated in a brass casing. The acoustic sensors were glued directly to the sample through pre-drilled holes in the jacket and detected  motion normal to the surface of the sample. 

During each test, unamplified signals from the 8 acoustic sensors were relayed to a 16 bit digital oscilloscope and recorded at 10 MHz sampling rate in a triggered mode. Unamplified waveforms were stored in blocks of 409.6 $\mu s$ (with an overlap of 102.4 $\mu s$) as long as the output voltage of one sensor exceeded a predefined threshold (20 mV). In parallel, signals from the 8 sensors were amplified at 45 dB before being relayed to a second 16-bit digital oscilloscope and continuously recorded at 10 MHz sampling rate. The purpose of recording the sensor signals at two different gains was to capture the large amplitude acoustic emissions generated by the stick-slip as well as the microseismicity (i.e. small amplitude AEs) before and after stick-slip.
AEs were searched within the continuous waveforms with a simple amplitude threshold algorithm: continuous waveforms were scanned with a $406.9$ $\mu s$ (i.e. 4096 samples at a sampling rate of 10 MHz) long moving time window and the acoustic waveforms were stored if any amplitude of the acoustic signal exceeded the predefined threshold on 2 or more channels. For each channel, the amplitude threshold was set to 1.5 times the noise level which was purposefully low in order to capture a maximum number of AEs and to build complete AE-waveforms catalogs. AE detection results were then double checked by hand in order to eliminate false detections. More details on the acoustic recording set-up and processing can be found in \citet{brantut2011damage} and \citet{passelegue2017influence}. 

\section{Methodology}
\subsection{Fault stress and fault slip calculation}
Axial displacement $D$ recorded at the top of the axial piston is equal to the sum of the fault displacement $\delta$ projected along the vertical axis and the elastic response $X$ of the sample and piston column such that: 

\begin{equation}
   D = \delta.cos(\theta)+X
\end{equation}
with  $\theta$ the angle between the normal of the fault plane and the axial stress $\sigma_{1}$. $X$ corresponds to the elastic response  of the combined system (sample+piston column) and is given by the ratio between the applied load $L$ and the apparent stiffness $k_a$ of the system:

\begin{equation}
   X = \frac{L}{k_{a}}
\end{equation}
Therefore, fault slip $\delta$ can simply be calculated as follows:
\begin{equation}
   \delta = \frac{D-L/k_{a}}{cos(\theta)}
\end{equation}
In practice, we made the assumption that the fault was fully locked (i.e, $\delta$ = 0) at the beginning of each stick-slip cycle. $k_{a}$ was determined from the slope of the linear elastic relationship between axial stress and axial displacement ($k_{a} = L/D$), at 25$\%$ of peak stress during the inter stick-slip event period. Since $k_{a}$ is susceptible to change with cumulative displacement, its value was updated at the beginning of each stick-slip cycle. Shear stress ($\tau$) and normal stress ($\sigma_{n}$) acting on the fault were then derived from Mohr circle equations:

\begin{equation}
   \tau = \frac{(\sigma_{1}-Pc)}{2}sin(2\theta)
\end{equation}
\begin{equation}
   \sigma_{n} =  \frac{(\sigma_{1}+Pc)}{2}+\frac{(\sigma_{1}-Pc)}{2}cos(2\theta)
\end{equation}
with $\sigma_{1}$ the axial stress and $Pc$ the confining pressure. Finally, friction is simply given by $\mu=\tau/\sigma_{n}$. 

\subsection{AE and Stick-Slip Event (SSE) hypocenter location}
For each stick-slip cycle, premonitory AEs and SSE-hypocenter were located by using a standard grid search method. The grid search analysis was restricted to a 2-D plane matching the elliptical saw-cut fault. The 2-D source location on the fault plane is given by the minimum of the L2 norm of the misfit, $M_{t}(x,y)$, between observed and theoretical P-wave arrival time differences ($\Delta t^{obs}_{i,j}$, $\Delta t^{t}_{i,j}$ respectively) for each pair of sensors $(i,j)$. For each grid node, $M_{t}$ takes the form:
\begin{equation}
    M_{t}(x,y)=\frac{\sum_{i=1}^{n}\sqrt{\sum_{j=1}^{n-1}(\Delta t_{i,j}^{obs} - \Delta t_{i,j}^{t})^{2}}}{n(n-1)}
\end{equation}
with $n$ the number of sensors, $(i,j)$ the pair of stations and $(x,y)$ the cartesian coordinates of the grid node. Theoretical arrival times were calculated using an isotropic P-wave velocity $c_{p}$ model. P-wave arrival times were first picked automatically using an Akaike Information Criterion (AIC) picking algorithm and then manually checked. Only AEs and SSE-hypocenters with location uncertainty lower than $\sim$ 2 mm (0.3 $\mu s$ time residual) were retained in our location maps. 
 
\subsection{Acoustic sensor calibration}
AE waveforms $S(t)$ may result from the convolution of three different terms:
\begin{equation}
    S(t)= f(t) * G(t) * R(t)
\end{equation}
where $f(t)$ is the source term, $G(t)$ is the impulsive response of the medium (or so-called Green's function) and $R(t)$ is the sensor response (i.e, the instrumental response). The purpose of calibrating an acoustic sensor is to characterize $R(t)$ which defines the response of the sensor to a mechanical input quantity (stress, displacement, velocity of acceleration).

The AE sensors used in this study were calibrated prior to the experiments, using a high-frequency laser interferometer. Figure 2a. displays a photograph and a schematic of the calibration set-up: an industrial (V109-rm or M110-sm, Olympus) P-wave transducer was positioned at the center of the saw-cut surface of the metagabbro sample and excited with a pick to pick 200 V sinusoidal wave. The generated wavefield was recorded on the opposite side of the sample, as particle velocity normal to the surface, with a laser interferometer (VibroOne, Polytec company) with a gain of 400 mm/s/V and a flat frequency response up to 2.5 MHz. One of the experimental (home-made) AE sensor was then glued to the sample at the location sampled by the laser. The transmitting transducer was similarly excited and surface vibrations were recorded again with the AE sensor. The instrumental response of the AE sensor $R_{s}(\omega)$ (V/mm/s) was then obtained in the frequency domain, in the range DC to 2.5 MHz, by simple deconvolution of the waveform recorded by the AE sensor $S_{a}(\omega)$ out of the waveform recorded by the laser $S_{v}(\omega)$:
\begin{equation}
    R_{s}(\omega)=\frac{S_{a}(\omega)}{S_{v}(\omega)}
\end{equation}
We explored calibration results reliability by varying (i) the type of transmitting transducer and (ii) the frequency of the sinusoidal voltage applied to the transmitting transducer (500 kHz, 1 MHz and 2 MHz). Transmitting transducers were of two types, namely V109-rm and M110-sm designed by Olympus company. These two transducers differ in size and shape but share similar characteristics: they produce longitudinal (P-)waves of 5 MHz central frequency. V109-rm has an L-shape and a nominal element size of 13 mm and M110-sm has a straight shape and a nominal element size of 6 mm. Calibration results are summarized in figure 2b. For a specific transmitter, we found that the excitation frequency had a negligible effect on the sensor's response (Figure 2b, two top panels). Regardless of the type of transmitter, the transducer's response was non linear with a clear resonance band between $\sim$ 1.2 MHz and 2.2 MHz. We found that, except for the width of the resonance band, the type of transmitter had little impact on the shape and amplitude of sensor sensitivity (figure 2b bottom). Because we expected AE source size to be of the order of a few mm, we opted to use the instrumental response derived from the calibration set-up with the smallest source, i.e. M110-sm. An important limitation of our calibration procedure is that it is probable that the instrumental response determined under atmospheric pressure might differ to the one under experimental pressure conditions. In particular, we expect the sensitivity of the sensor to increase, particularly at high frequency, under high pressure conditions, due to increasing sensor/rock coupling. At the same time, it is probable that sensor resonance is damped by the high pressure oil used a confining medium. These effects have not been investigated.

For each AE recorded during the experiment, the velocity spectrum $\Omega_{v}(\omega)$ was calculated by simple deconvolution of the AE waveform spectrum $S_{a}(\omega)$ with the instrumental response $R_{a}(\omega)$ such that:
\begin{equation}
\Omega_{v}(\omega)=\frac{S_{a}(\omega)}{R_{a}(\omega)}
\end{equation}

Ultimately, the displacement spectrum, $\Omega_{d}(\omega)$, from which AE source parameters were derived (see next section), was obtained by integration in the frequency domain:
\begin{equation}
    \Omega_{d}(\omega)=\frac{\Omega_{v}(\omega)}{2\pi \omega}
\end{equation}

\subsection{Inversion of AE source parameters}
For each AE, absolute moment magnitude $M_{w}$, corner frequency $f_{c}$, static stress-drop $\Delta \sigma$ and source radius $r$ were derived from the stacked  S-wave displacement spectrum. Single-sensor displacement spectra were obtained from a 27.5 $\mu s$ long time-window starting 2.5 $\mu s$  before the S-wave arrival. The first 2.5 $\mu s$ were multiplied to a ramp function to minimize the effect of P-wave energy on the results, the resulting signal was rescaled to a 50 $\mu s$ long time window centered on S-wave arrival and multiplied to a Hann window. Note that theoretical S-wave arrival was calculated according to the source-receiver distance and the S-wave velocity of the rock sample.

S-wave spectra $\Omega_{s}$ were modelled using a Brune source model with an attenuation term such as:
\begin{equation}
    \Omega_{s}(f) =\frac{\Omega_{0}}{1+(f/f_{c})^2}\exp(-\pi f t/Q)
\end{equation}
where $\Omega_{0}$ is the long-period spectral plateau, $t$ is the S-wave travel time averaged over all the sensors, $Q$ the attenuation factor and $f_{c}$ the corner frequency. $f_{c}$, $\Omega_{0}$, and $Q$ were estimated by grid search in the ranges of 100 kHz to 2.5 MHz, $10^{-18}$ to $10^{-15}$ $m.s^{-1}$ and 30 to 50 respectively. The range of values for Q was selected according to values found in literature \citep{goldberg1992acoustic,liu1997stress,yoshimitsu2014magnitude} and was intentionally narrow to avoid significant trade-offs between Q and $f_{c}$ that both control the high-frequency spectral decay.

The seismic moment $M_{0}$ was computed from $\Omega_{0}$ according to:
\begin{equation}
    M_{0}=\frac{4\pi}{\Lambda_{\theta,\phi}}\rho c_{s}^3R\Omega_{0}
\end{equation}
 where $\rho$ is the bulk density, $c_{s}$ the S-wave velocity, $R$ the average distance between the source and the sensor and $\Lambda_{\theta,\phi}$ the averaged S-wave radiation pattern (0.63, \citet{aki2002quantitative}). From $M_{0}$, we get the absolute moment magnitude as:
\begin{equation}
M_{w}=(\log_{10}(M_{0})-9.1)/1.5
\end{equation}
We used the circular crack model of Madariaga \citep{madariaga1976dynamics} to estimate the radius $r$ from $f_{c}$ such as:
\begin{equation}
   r=\frac{0.21.C_{S}}{f_{c}}
\end{equation}
Finally, the stress drop $\Delta\sigma$ was calculated from $M_{0}$ and $r$ as \citep{eshelby1957determination}:
\begin{equation}
   \Delta\sigma=\frac{7M_{0}}{16r^{3}}
\end{equation}
 Figure 2c displays an example of fitted displacement spectra for two AEs of moment magnitudes $Mw$ -7.7 and $Mw$ -8.6. The estimated corner frequencies are 0.88 MHz and 1.5 MHz, source radii are 0.8 mm and 0.45 mm and static stress drops are 0.75 MPa and 3.35 MPa respectively.  

 \subsection{Estimation of b-value}

AE frequency-magnitude distributions were modelled with a tapered Gutenberg-Richter (GR) distribution. The tapered Gutenberg-Richter relation has an additional exponential term with respect to the classical GR and is commonly expressed in terms of seismic moment $M$:

\begin{equation}
  N(M)=N_{t}.(M_{t}/M)^{\beta}.\exp(\frac{M_{t}-M}{M_{c}})  
\end{equation}
with $N(M)$ the number of events with seismic moment larger than $M$, $N_{t}$ the number of events with seismic moment larger than the completeness seismic moment $M_{t}$, $M_{c}$ is the 'corner' seismic moment that controls the distribution in the upper range of $M$ and $\beta$ is the exponent of the distribution which is equal to 2/3 of the b-value ($\beta$ = $\frac{2}{3}$$b$). $\beta$ and $M_{c}$ were estimated by grid search analysis (minimizing the misfit between model and observation) and b-value was obtained from $\beta$.

\section{Experimental results}
\subsection{Mechanical data}
Figure 3 displays friction, cumulative fault slip (stiffness corrected) and AE rate as a function of time during three experiments performed at Pc = 30, 45 and 60 MPa respectively. Experiments were stopped after 8 $mm$ of cumulative displacement and a total of 55 SSEs, 29 SSEs and 13 SSEs were produced at Pc = 30, 45 and 60 MPa respectively. The fault response to mechanical (constant far-field displacement) loading was similar at Pc = 30 and 45 MPa (figure 3, top 2 panels). The first SSE occurred when the static friction coefficient reached $\sim$ 0.5. Then static (peak) friction coefficient and co-seismic stress-drop continuously increased up to the end of the experiments. Static (peak) friction increased from 0.5 to 0.7 and  frictional stress-drop increased from $\sim 0.05$ to $\sim 0.1$ for both confining pressures. At Pc = 60 MPa, the first SSE occurred when the static friction coefficient reached $\sim$ 0.65. Unlike the other two experiments, static friction did not continuously increase with cumulative slip but rather fluctuated between $\sim$ 0.65 and 0.72. Co-seismic frictional stress drop were also in the order of $\sim 0.1$. One interesting (and yet often disregarded) observation, is that while saw-cut faults were locked at the early stage of loading, they rapidly unlocked and began to slip, regardless of confining pressure. Indeed, stable fault slip started as early as the friction coefficient reached $\sim$ 0.2-0.25 as illustrated by the change in the shear-stress and slip versus time slopes and a first SSE was generally required for the fault to enter a 'pure' stick-slip behavior, i.e. deform solely elastically during inter-event loading.

Three common features between the experiments are that AE activity remained concentrated in the last seconds prior to failure (as illustrated by the bursts of AE activity, figure 3) and that the fault ruptured several times before generating AEs (low or no AE activity during the first couple of SSEs). It is interesting to point out that these experiments never produced aftershocks, even upon looking carefully within the continuous wafevorms, which probably highlight the fact that the stress drop (and hence slip) had been rather homogeneous on the fault during the propagation of the mainshock. So all the detected AEs here can be considered as foreshocks.  However, from a general point of view, we observed a great degree of variability in the AE foreshock activity, which tend to increase with increasing confining pressure conditions. For instance, while some SSEs produced no AEs at Pc = 60 MPa, the highest rate of foreshock triggering (39 AEs/s) was also recorded during this experiment. In contrast, foreshock activity at 30 MPa fell rapidly into a more predictable behavior characterized by a systematic AE rate increase prior to mainshock propagation. 

\subsection{AE distribution}
Figures 4a, 4b and 4c display the number of AEs (dark color histograms), and the AE moment release (light color histograms) per stick-slip cycle at Pc = 30, 45 and 60 MPa respectively. In total, 905, 380 and 185 AEs were recorded at Pc = 30, 45 and 60 MPa respectively, which equates to an average number of AEs per SSE of $\sim$ 17, 13 and 14 respectively. The maximum number of AEs per SSE that were recorded is 48, 31 and 46 at Pc = 30, 45 and 60 MPa respectively. We estimated that the maximum AE moment was of 0.8 N.m at Pc = 30 MPa and 0.18 N.m at  Pc = 45 and 60 MPa. We observe here in more details that both the number of AEs and their cumulative moment remained quite variable during the experiments, although an increasing trend with cumulative slip can be observed, albeit less significant at Pc = 60 MPa. The origin of this variability seemed uncorrelated either to the prior or the coming stress drop magnitude and will remain beyond the scope of the present study. 

AE moment magnitude and AE seismic moment are displayed as a function of corner frequency in Figure 4d. Errorbars for corner frequency and moment magnitude are indicated in light gray. AE moment magnitudes range between $\sim$ -8.6 and $\sim$ -7. For the large majority of the microseismicity, corner frequencies and stress-drops range from 300 kHz (source size $\sim$ 4 mm) to 1.5 MHz (source size $\sim$ 0.5 mm) and from 0.1 to 10 MPa with an average value for the stress-drop of 1 MPa, 0.88 MPa and 0.68 MPa at Pc = 30, 45 and 60 MPa respectively. A common observable is that larger AEs have larger stress-drops. The origin of this trend may result from the fact that, as will be seen in the next sections, larger AEs tend to occur closer to mainshock propagation, and, in such way, could in effect have larger stress drops as the fault surface approaches criticality and slip rate (and thus stress release) accelerates. Note that the possible bias mentioned above in our calibration procedure - the fact that high-frequency waves may propagate better under confinement than under room pressure, because both the attenuation of the rock medium will decrease as pre-existing cracks will close and so will the contact between the transducer and the rock surface - would tend to act in the opposite way. Indeed, one would have expected high-frequency amplitudes to be larger under confinement than under room pressure conditions, and in consequence, to overestimate the stress-drops of small magnitude (high-corner frequency) AEs. However, the evolution of transducer resonance may also have an effect, difficult to intuit, under confinement, another alternative being a frequency dependent quality factor. 

\subsection{Foreshock and mainshock hypocenter locations}
Photographs of the saw-cut faults after the experiments (left) with AE (circle symbols, middle) and SSE hypocenter (star symbols, right) locations are shown in Figure 5. Colorscale refers to SSE index and AE source sizes match source radii obtained by inversion. For a $M_{w}$ = -7  event and for a $M_{w}$ = -8 event, source size is typically of the order of 3 mm and 1 mm respectively. The largest amount of gouge was produced at the lowest stress condition (Pc = 30 MPa). Wear product (white patterns, left photograph) aggregated into millimetre scale patches whose elongated shapes highlight the direction of sliding. Gouge clusters are fairly homogeneously distributed on the fault plane, except near the fault edges where very little gouge was produced, which is a common feature to the three experiments. For the intermediate stress condition, gouge particles also gathered in elongated clusters in the sliding direction, however their spatial distribution is significantly more heterogeneous as illustrated by the dark spots where the fault remained almost intact (no wear production). It should be noted that post experiment, the two fault surfaces were always found symmetrical, so these spots without gouge are not due to possible gouge removal when the two pieces were separated after the tests. In contrast, gouge distribution is more homogeneous at the highest stress condition (Pc = 60 MPa), during which wear material did not form clusters and sliding direction is hard to guess at that scale.

Although it is clear from the experiments conducted at Pc = 30 MPa and Pc = 45 MPa that AE locations correlate rather well with gouge distribution, gouge product does not necessarily implies AE activity. This last observable is well illustrated by the fault surface at Pc = 60 MPa where a significant surface of the fault, while covered with gouge, did not produce AEs. SSE hypocenter locations migrated during the experiments, which is best illustrated at the two lowest stress conditions. SSE hypocenters initially located in the middle of the fault and then propagated to both ends of the fault at Pc = 30 MPa. SSE hypocenters first located on the lower right edge of the fault and then migrated to the lower left edge at Pc = 45 MPa.  Regardless of the stress conditions, AE and SSE hypocenter locations and migrations match fairly well with one another. However, a noticeable mismatch between AE and SSE hypocenter locations is observed at the end of the experiment conducted at Pc = 30 MPa.  

\subsection{Microstructural analysis}
Fault surfaces were observed under scanning electron microscopy (SEM) after the experiments (Figure 6). From the images at the finest scale (Figures 6a, 6c and 6e), we evaluate the range of gouge particle sizes to be between less than 1 $\mu m$ and few $\mu m$. Gouge particles cover asperity-like topographic heights, with size of the order of few tens of $\mu m$ (Figures 6b, 6d and 6f). While it is difficult to infer slip direction from geometric patterns at Pc = 30 MPa, fault surfaces at Pc = 45 MPa and 60 MPa show evidence of thermally induced plastic deformation and melting processes, respectively, both of which captured the slip direction (Figures 6c-f). At Pc = 45 MPa, fault surface presents compacted and flatten microstructures (Figure 6c), aligned with the slip direction, which evidences that, locally at least, the fault surface temperature has nearly reached the melting point ($\approx$ 1200$^{\circ}$). Finally, the elongated and stretched stringy microstructures observed at Pc = 60 MPa (Figure 6e) is a robust proof of asperity melting of the fault surface during slip. The micro-crack (Figure 6d) perpendicular to slip direction that crosses the residual melt is likely due either to rapid cooling following melting or to co-seismic damage. 

\section{Discussion}
\subsection{Nucleation phase dynamics}
Figures 7a, b and c compare the along fault displacement (blue curves), the along fault velocity (red curves), the cumulative number of AEs (black curves) and the cumulative AE moment release (dashed curves) with respect to time to failure at Pc = 30, 45 and 60 MPa  respectively. The grey shaded area indicates the range of uncertainty for the cumulative AE moment release. Each quantity has been normalized by its maximum value at the time of failure and all the premonitory sequences have been stacked to highlight the general trend. Consistently with previous experimental studies \citep{mclaskey2014preslip,passelegue2017influence,yamashita2021two}, we observed fault displacement preceding failure. However, although fault slip is required to generate foreshocks, both the number of foreshocks and their cumulative moment appear to correlate with slip velocity rather than slip itself. This is particularly well illustrated in the last seconds prior to failure during which cumulative moment release and fault slip velocity almost collapse. 

Nevertheless, there are noticeable differences between the experiments. The cumulative AE moment release exhibits the smoothest behavior at Pc = 30 MPa. Seismic energy is continuously radiated from the fault but in a delayed fashion with respect to slip velocity. For instance, between about -15s and -5s,  slip velocity increases while the AE moment release remains low. These features can also be retrieved for the experiments conducted at Pc = 45 MPa, during which the moment release is delayed relative to slip velocity and intensifies only once the fault accelerates. In the same way than during the experiment conducted at Pc = 30 MPa, the slip velocity increases linearly before accelerating (between about -5 and -2s prior to failure). In summary, with increasing confining pressure, the slip velocity and the AE moment release also increase later, i.e. closer to the stick-slip instabilities, but then accelerate faster.

The picture depicted by the experiment conducted at Pc = 60 MPa is somewhat different. Although we observe a clear correlation between the fault slip velocity and the cumulative moment release, the seismic energy is not released continuously, but in bursts. For instance, the two largest AEs that were recorded ($Mw$ $>$ -6.9) occurred about -17s and -5s prior to failure, while the fault had not accelerated yet. This case is not limited to the experiment conducted at Pc = 60 MPa and also occurred at Pc = 30 MPa and Pc = 45 MPa. Even at the small scale of the experiments presented here, foreshocks may occur in bursts without apparent external forcing such as slip acceleration. This might reflect the brittle failure of small patches where residual stress accumulated. Also, the stacking procedure inherently smooths the variability of precursory AE sequences.  It is likely that bursts of AE activity would have been smoothed if a larger number of AE sequences would have been stacked together at Pc = 60 MPa.

\subsection{Variability and fault maturation}
The cumulative precursory foreshock activity per SSE is plotted versus time to failure at Pc = 30, 45 and 60 MPa figures 7d, e and f respectively. The black curve corresponds to the average sequence, while colored curves represent individual foreshock sequences, with the colorscale referring to the SSE index. Note that for visual inspection, not all AE sequences are shown at Pc = 30 MPa (Figure 7d) and at Pc = 45 MPa (Figure 7e). The experiment conducted at Pc = 30 MPa gives the clearest example of what one would call "fault maturation" (Figure 7d). At the early stage of the experiment, most of the foreshocks occurred within seconds to failure, but with successive ruptures, precursory AE activity increased in number and occurred earlier during loading. Both at Pc = 30 MPa and Pc = 45 MPa, the number of foreshocks prior to failure only started to significantly increase after 10 stick-slip cycles. 

Summing all AE sequences results in a smooth increase of the cumulative number of foreshocks as previously described. A noticeable difference lies in the absence of foreshocks early during loading at Pc = 60 MPa. During this experiment, foreshocks occurred later, which results in a sharper acceleration of the cumulative number of foreshocks towards failure. On the other hand, the experiment conducted at Pc = 60 MPa is the only one for which the first foreshocks have released a large amount of seismic energy early in the sequence, with respect to the ones that followed (Figure 7c). 

\subsection{Inverse Omori-law}
When averaged over numerous foreshock sequences, it is known that the foreshock rate $N(t)$ increases as an inverse power law of the time to the mainshock \citep{jones1979some,shearerearthquake} which, by analogy with the direct Omori's law, can be expressed as:
\begin{equation}
N(t)=\frac{K}{(c+\Delta t)^{p}}
\end{equation}
where $K$ is the foreshock productivity, $c$ and $p$ are empirical constants and $\Delta t$ is the time to mainshock (or failure). Figures 7d, e and f show the stacked cumulative number of foreshocks $N_{a}(t)$ in the last 40 seconds prior to failure at Pc = 30, 45 and 60 MPa respectively. This allows us to highlight the smooth shape of the cumulative total number of AEs and to compare between the experiments the average number of precursory AEs during individual sequence. 

The parameters $p$ and $c$ were searched in the range [0.1-3] with a step of 0.01. We made the choice to link $K$ to $c$ and $p$ such as $K = N_{f}.(c^{p})$ where $N_{f}$ is the average cumulative number of AEs at the time of failure. This ensures that the average cumulative number of AEs at the time of failure equals $N_{f}$. The best fits were obtained for $c$ = 2.39 $\pm$ 0.3s and $p$ = 1.31 $\pm$ 0.08, $c$ = 0.6 $\pm$ 0.25s and $p$ = 0.79 $\pm$ 0.1 and $c$ = 0.24$\pm$ 0.09 s and $p$ = 0.82 $\pm$ 0.05 at Pc = 30, 45 and 60 MPa respectively. 

In summary, we find that both $p$ and $c$ decreases  with increasing normal stress (confining pressure). According to (17) and using the best set of parameters obtained for $c$ and $p$, we find that the average AE rate is about 5 times larger at Pc = 60 MPa compared with Pc = 30 MPa and about two times larger at Pc = 45 MPa compared with Pc = 30 MPa at the time of failure. This correlates well with the fault slip velocity. If we compare with the average fault slip velocity in the last ten milliseconds we find that the fault slip slip velocity is about four times larger at Pc = 60 MPa (about 4 $\mu m/s$) compared with Pc = 30 MPa and about three times larger at Pc = 45 MPa compared with Pc = 30 MPa. Given the good correlation that we found between fault slip velocity and AE cumulative number (Figures 7a, b and c), we posit that AE rate is primarily controlled by fault slip rate. However, it should be noted that this is only valid on average since precursory AE sequences exhibit variable behaviors with respect to each other.

We note two important factors that may bias our estimations of $p$ and $c$, (i) we expressed $K$ as a function of $c$ and $p$ and (ii) we may have missed a significant number of AEs close to failure, either because of AEs occuring at the same moment and at the same location or because of small AEs that would be hidden by bigger ones. The most common way to estimate $K$, $c$ and $p$ is to use the maximum likelihood method \citep{ogata1983estimation} which also quantifies the interdependence of $K$, $c$ and $p$. Since we have expressed $K$ as a function of $c$ and $p$ in the same way for each experiment and that $N_{f}$ do not differ much ($N_{f}$ equals 17, 13 and 14 at Pc = 30, 45 and 60 MPa respectively) we believe that linking $K$ to $c$ and $p$ does not preclude interpreting the results relative to each other. Finally, AE catalogs magnitude completeness $M_{c}$ in the last seconds prior to failure does not vary significantly with stress conditions ($M_{c}$ $\approx$ -8.4). Therefore, missed AEs are likely to influence the absolutes estimates of $p$ and $c$ but not their relative values.

\subsection{Scaling with nucleation size}
When derived from ETAS (Episodic Type AfterShock Sequence) models \citep{helmstetter2003foreshocks}, the value of $p$ is universal and close to unity. In ETAS models, the inverse Omori's law for foreshocks stems from the combination of the direct Omori's law (i.e., any earthquake triggers its owns aftershocks) and the triggering of earthquakes in "cascade" mode due to stress (static or dynamic) transfer. 

Figure 8a displays best fits for inverse Omori-law (dashed curves) super-imposed to stacked sequences fixing $p=1$. In this case, the best values of $c$ are $c$ = 1.41, 0.92 and 0.35s at Pc = 30, 45 and 60 MPa respectively, i.e. a clear linear decrease of $c$ with increasing normal stress at the onset of failure (inset in Figure 8a). Such observation is reminiscent of several experimental studies that have already proposed an inverse dependence of characteristic 'nucleation' time with stress \citep{latour2013characterization, Gvirtzman2021nucleation}, while the inverse dependence of the $c$-value on stress conditions has also been inferred from natural aftershock data  \citep{narteau2002temporal,narteau2009common}.

A possible interpretation consistent with the above observation is if the characteristic nucleation time scales with the critical nucleation length $L_{c}$. For a linear slip weakening friction law \citep{ida1972cohesive,campillo1997initiation,uenishi2003universal}, $L_{c}$ is defined as:
\begin{equation}
    L_{c} = \beta\frac{\mu D_{c}}{\sigma_{n}(f_{s}-f_{d})} 
\end{equation}
where $\mu$ is the shear modulus of the rock sample, $D_{c}$ is the critical slip distance, $\sigma_{n}$ is the normal stress acting onto the fault, $f_{s}$ and $f_{d}$  are the static and the dynamic friction coefficients respectively and $\beta$ is a non-dimensional shape factor coefficient ($\approx$ 1.158). Consequently, $L_{c}$ is expected to decrease with increasing normal stress, as supported by stick-slip experiments on plastic polymers \citep{latour2013characterization}.

Taking $D_{c}$ equals to the average pre-slip ($\approx$ 5 $\mu m$), $\sigma_{n}$ equals to the average normal stress at the time of failure ($\approx$ 50, 70 and 100 MPa at Pc = 30, 45 and 60 MPa respectively) and $\mu$ = 35 GPa, we find reasonable estimates (ie. comparable or smaller than the fault total length size) of $L_{c}$ equal to $L_{c}$ = 80, 58 and 40 mm, respectively, only if $(f_{s}-f_{d})$ $\geq$ 0.05, which, in turn, is a rather unreasonably estimate of friction drop during the nucleation phase. However,  $\sigma_{n}$ is likely to be underestimated since the real contact area of the fault is less than the apparent fault surface area. $D_{c}$ is also an upper bound since it equals to the total displacement along the fault from the beginning to the end of loading. In addition, if $L_{c}$ was larger than the size of the fault for all experiments, no frictional instability would occur. On the other hand,  the fact that AEs tend to be distributed over the entire fault surface (Figure 5) suggests that $L_c$ must be probably comparable to the experimental fault size. This is also in agreement with the total area covered by foreshocks which seems to decrease with increasing stress conditions (Figure 5).

An alternative is to calculate either of the critical nucleation lengths ($L_{a-b}$, $L_{b}$) of the Rate and State friction law, defined as \citep{ rubin2005earthquake}:
\begin{equation}
    L_{a-b} = \frac{\mu D_{c}}{(b-a)\sigma_{n}};
    L_{b} = \frac{\mu D_{c}}{b\sigma_{n}} 
\end{equation}
where $a$, $b$ and $D_c$ are constitutive parameters. For Indian metagabbro, $a$, $b$ and $D_c$ were precisely determined under slow loading conditions ($\sim$ 10 $\mu m/s$), but albeit relatively low normal stress to be  $a=0.005$, $b=0.009$ and $D_c$ $\sim$ 1 $\mu m$ \citep{urata2018apparent}. Under our experimental conditions, this would correspond to nucleation lengths $L_{a-b}=175; 125; 87$ mm and $L_{b}= 78; 55; 38$ mm  at Pc = 30, 45 and 60 MPa respectively. In conclusion, we find $L_b$ to be the most consistent nucleation length estimate with our experimental observations, which is expected in the case of strong rate weakening \citep{viesca2016stable}.  

Now, scaling both the foreshock productivity and the time to failure with the nucleation length $L_b$, we find that foreshock sequences collapse onto a single inverse Omori master curve (Figure 8b). One can try to intuit what the scaled productivity (number of foreshocks/m) and the scaled time to rupture (s/m) represent physically. The latter may be interpreted as the inverse of fault slip velocity. It is interesting to point that the acceleration happens for a slowness of the order of $\sim$ 100 s/m, i.e. a corresponding slip velocity in the range of cm/s, which is the typical slip-velocity at which thermal weakening processes start to be activated at the laboratory scale \citep{rice2006heating,di2011fault, goldsby2011flash}. This is also in line with previous experimental works on stick-slip dynamics \citep{latour2013characterization, passelegue2016dynamic}. The scaled foreshock productivity may in turn reflect the fault motion inside the nucleation zone, i.e. the number of foreshock produced per amount of slip or advancement of the detachment front, which might be controlled by fault mechanical properties such as roughness for instance and, of course, loading conditions.

In summary, when taken individually, foreshock sequences are characterized by high variability which , as suggested theoretically \citep{lebihain2021earthquake,schar2021nucleation} and experimentally \citep{gounon2022rupture}, is the manifestation of complex rupture nucleations on an hetererogeneous fault interface. Because stacking the foreshocks sequences smooths this variability, our interpretation of the scaling on the inverse Omori law with normal stress/nucleation size, is thus that stacked foreshock sequences do indeed reflect the homogenized nucleation of the mainshock itself as conceptualised by Ohnaka's model \citep{ohnaka1992earthquake}. From a qualitative perspective, if foreshocks are driven by the nucleation phase of the upcoming mainshock, their temporal distribution should satisfy a universal temporal distribution, which corresponds to the dynamics of the nucleation phase itself.

\subsection{Gutenberg-Richter $b$-value temporal evolution}
We now look at the temporal evolution of the $b$-value of the stacked precursory AEs prior to failure (Figures 9a, b and c) at Pc = 30, 45 and 60 MPa respectively. b-values were estimated within a moving window (with a tapered Gutenberg-Richter model, see section 3.5) containing 100 AEs at Pc = 30, 45 MPa and 50 AEs Pc = 60 MPa. Shaded areas indicate the 90$\%$ confidence intervals.

At Pc = 30 MPa we estimate the $b$-value to be about 0.9 in the $\sim$ 10 seconds prior to failure. The $b$-value then dropped  rapidly to reach an almost constant level $\sim 0.4$ a at the time to failure of $t\sim2s$, which corresponds roughly to the $c$-value (dashed line) of the inverse Omori-law. The same is observed at Pc = 45 MPa, during which the $b$-value was close to 1 up to $\sim$ 10s prior to failure and then abruptly dropped to reach an almost constant level $\sim 0.5$ at a time to failure of $t\sim1$s, which again corresponds roughly to the $c-$value  (dashed line) of the inverse Omori-law. For both experiments, the $b$-value remained constant and equal $\sim$ 0.4-0.5 until failure. Unlike the other two experiments the $b$-value for the experiment conducted at Pc = 60 MPa is initially low, close to $0.6$, increases up to $0.8$ and then decreases again to reach a value close to $0.5$ in the last tenths of a second before rupture. However, the temporal evolution of the $b$-value prior to failure is less reliable during this experiment  for at least two reasons: i) the number of foreshocks was considerably lower (185, compared to 905 and 380 at Pc = 30 and 45 MPa respectively) and ii) close to 90 $\%$ of these were recorded in the last 3 seconds prior to failure which lowers considerably the temporal resolution of $b$-value variations in the early stages of the stick-slip cycles. In comparison about 30 $\%$ and $25 \%$ of the total number of AEs were generated before entering the last 3 seconds prior to failure at Pc = 30 and 45 MPa respectively. 

\citet{aki1981probabilistic} proposed a model where the fractal dimension of the fault plane is equal to $\sim b/2$. In which case, a $b$-value of $1$ corresponds to productivity on a plane, while in his model, a $b$-value of $0.5$ would correspond to fault lines filling up a plane. Based on the above observations, we can thus interpret the drop in $b$-value from $\sim 1$ far away in time from rupture, to $\sim0.5$ at the onset of slip acceleration as a transition from foreshocks being produced on the entire fault plane, to foreshocks being the result of an accelerating slip front due to the rapid weakening of the fault interface close to failure. Temporal variations in $b$-value prior to failure have also been documented during fracture experiments conducted on intact rock samples \citep{scholz1968frequency,lockner1991quasi} and during rock friction experiments \citep{goebel2012identifying,kwiatek2014seismic,riviere2018evolution}. Fracture experiments on intact samples show that $b$-value and differential stress are anti-correlated, which takes its origin in the formation and the coalescence of microfractures. Such a process causes a large number of AEs to be generated and a smooth and accelerating drop of $b$-value up to the time of failure. Decrease in $b$-value towards failure has also been documented preceding large subduction earthquakes \citep{suyehiro1966difference,enescu2001some,nanjo2012decade,tormann2015randomness}. However, foreshocks that precede large earthquakes occur on time scales from hours to years. Long term variations of $b$-value are usually attributed to stress accumulation or partial stress release while short term variations are related to the mainshock nucleation. 

\subsection{Foreshock migration}
We now look at the evolution of the spatial distribution of foreshocks towards failure (Figures 9d, e and f). Shaded areas indicate the $\pm$ one standard deviation intervals. In what follows, "nucleation" refers to the mainshock hypocenter on the fault surface, which was determined using first P-wave arrival times (Figure 5, right panels). Mainshocks whose nucleation sites were poorly constrained (less than about 2-3 mm) were not taken into account in the following analysis. Note that, foreshocks located with more than 0.3 $\mu s$ travel time residuals (about 2-3 mm of location accuracy) were also disregarded.

Figures 9d, e and f display the average distance to SSE hypocenter of the stacked precursory AEs as a function of time to failure at Pc = 30, 45 and 60 MPa respectively. Average distance to SSE hypocenter was computed within a moving window containing 50 AEs at Pc = 30, 45 MPa and 25 AEs at Pc = 60 MPa. Note that decreasing the size of the window has no impact on the results other than to introduce high-frequency oscillations. At Pc = 30 MPa, we see that the average distance of the foreshock sequence continuously decreases when approaching failure,  to finally stabilize at an average of 25 mm to the eventual mainshock epicenters. The same is observed at Pc = 45 and 60 MPa, during which the average distance of the foreshock sequence continuously decreases when approaching failure,  to finally stabilize at an average distance of 20 and 15 mm respectively of the eventual mainshock epicenters. One should first note that, given the uncertainties, these values are compatible with our former estimates of $L_b/2$.

The spatial distribution of foreshocks yields relevant information about the way mainshocks initiate. In all experiments, we found that SSEs are always preceded by pre-slip acceleration phase. Moreover, we found that mainshocks do not necessarily nucleate where foreshocks concentrate, but rather at the edges of the areas where most of the precursory AE moment was released. Because of the drop in $b$-value discussed above, large precursory AEs, which tend to occur closer to failure, may promote a cascade-like process. Previous experimental studies \citep{mclaskey2014preslip,mclaskey2019earthquake, passelegue2017influence} proposed that pre-slip may sufficiently weaken fault strength to facilitate a small instability to grow large and eventually propagate over the entire fault. In such a scenario, precursory AE activity should migrate towards the mainshock epicenter in the last milliseconds prior to failure, as observed for the first time experimentally here. 

\subsection{Seismic coupling during nucleation}
We now compute the evolution of fault coupling, i.e. the ratio between the moment released by foreshocks (black curves in Figure 7) and that by fault slip (blue curves in Figure 7) as a function of time to failure. Fault coupling was computed within a moving window of duration equal to 1 $\%$ of the duration of the stacked precursory AE sequence. The moment released by fault slip is simply computed as $M_{0s}=G\pi ab\delta$, where $a$, $b$ are the long and short axis of the elliptical fault and $\delta$ the fault slip. This ratio is plotted as a function of time to rupture for the three experiments at Pc = 30, 45 and 60 MPa (Figures 9g, h, i respectively). Shaded areas display the range of uncertainty computed from foreshock magnitude uncertainties. In all three experiments, the cumulative moment release of foreshocks during nucleation represents only a very small percentage of the pre-seismic slip (Figures 9g, h and i). Yet, it continuously increases during nucleation, from a fraction close to zero at the beginning of nucleation, to $3\%$, $0.5\%$ and $0.2\%$ at the onset of failure  at Pc = 30, 45 and 60 MPa respectively. Moreover, one can observe that the coupling increases drastically during the phase of nucleation when the distance of foreshocks to the epicenter as well as the $b$-value have stabilized. Before that, some transient increases in coupling are observed, mainly due to foreshocks occurring in bursts. 

Our observations thus demonstrate that, at least in our experiments, the nucleation phase initiates as an almost fully aseismic process, and transitions, as time to failure approaches, towards a cascading process. In that interpretation, both the cascade and pre-slip models are not exclusive, as previously noted by \citet{mclaskey2019earthquake}. It is also interesting to note that the observed value of coupling at the onset of the mainshock decreases with increasing normal stress, which, again, is reminiscent of a shrinking of the nucleation size discussed above. In such case, the amount of fault area which radiates 'seismically' at the final stage of nucleation can be interpreted as a lower bound for the nucleation size. The values of $3\%$, $0.5\%$ and $0.2\%$ observed close to failure translate in a minimum nucleation length of $\sim10$, $\sim4$ and $\sim2.5$ mm, or, assuming a stress drop of 10 MPa, an equivalent moment magnitude of $-5.4$, $-6.2$ and $-6.6$ at Pc = 30, 45 and 60 MPa respectively, which is larger than the largest foreshocks detected ($M_{w} \sim -6.8$) in our experiments. Note that seismic coupling could not be resolved in the last tens of milliseconds (doubled headed arrows, figures 9g-i) due to insufficient temporal resolution ($\sim$ 0.1s), thus the aforementioned minimum nucleation lengths are likely to be underestimated. One can attempt to extrapolate seismic coupling using the linear portions of the curves in the last second prior to failure. By doing so, we obtain values of $\sim$ 6$\%$, 1.5$\%$ and 1.2$\%$ entering the last ten milliseconds prior to failure (i.e, $10^{-2}$s) at Pc = 30, 45 and 60 MPa, which equates to a minimum nucleation length of $\sim14$, $\sim7$ and $\sim6$ mm.

\section{Scaling laws and implications for natural earthquakes}
 \subsection{Pre-seismic moment and seismic coupling}
 Figure 10a compares the total foreshock moment release per mainshock $M_{0a}$ with the pre-seismic moment release $M_{0p}$. Figure 10b shows pre-seismic moment release as a function of co-seismic moment release. Our data (diamond symbols) are plotted together with the observations made by two previous experimental studies (\citet{passelegue2017influence,acosta2019precursory}, grey symbols). Pre-seismic moment release and co-seismic moment release were estimated according to $M_{0p,0c} = \mu D_{p,c} S $ with $\mu$ being the metagabbro shear modulus, $S$ the surface of the fault and $D_{p}$ and $D_{s}$ the pre-seismic slip and the co-seismic slip respectively. $D_{p}$ is the total macroscopic fault slip accumulated between two successive mainshocks and thus includes pre-slip related with nucleation and potential fault creep. Here after, we refer as to "seismic coupling" the ratio between the total moment AE release $M_{0a}$ and the pre-seismic moment release $M_{0p}$.

Seismic coupling ranges from about $10^{-6}$ ($10^{-4} \%$) to $10^{-3}$ (0.1 $\%$). Regardless of stress conditions, we find a power law between $M_{0a}$ and $M_{0p}$ of the type $M_{0a} \propto M_{0p}^{n}$. Although, we acknowledge that data range and data quality preclude a robust estimate of the power law exponent $n$, (3:5) seems a reasonable range of values (Figure 10a.). In the case of an isotropic expansion of a self-similar crack of  length $L$, the moment release inside the crack scales as $\Delta\tau L^{3}$ or, equivalently, as $\Delta\tau D_{i}^{3}$ \citep{madariaga1976dynamics} with $D_{i}$ the amount of slip inside the crack. Thus the latter scenario predicts that $M_{0a}$ goes as $M_{0p}^{3}$ (i.e. $n$ = 3) and is consistent with the interpretation made so far which is that mainshocks initiated as the emergence of an aseismically slipping fault patch that was driving precursory foreshock activity. The case $n$ $>$ 3 can be explained if foreshocks have stress-drops that are magnitude dependent, that is higher stress-drops for larger magnitudes. \citet{blanke2021stress} used the spectral ratio technique to precisely estimate AE stress drops and observed a significant increase in AE stress drop with AE size, which is a feature we also observe (Figure 4d).

\citet{acosta2019precursory} argued that the pre-seismic moment release $M_{0p}$ should scale with the co-seismic moment release $M_{0c}$. This scaling relationship is indeed expected if fracture energy increases as a power law of co-seismic displacement \citep{abercrombie2005can,ohnaka2013physics,passelegue2016dynamic} such as:
\begin{equation}
    G=\zeta u_{cos}^{\alpha}
\end{equation}
where $\zeta$ is a scaling pre-factor, $\alpha$ is the scaling power-law exponent and $u_{cos}$ is the co-seismic displacement. The following empirical scaling relation between $M_{0p}$ and $M_{0c}$ was proposed (indicated by the slope = 0.56, figure 10b):

\begin{equation}
    M_{0p} \propto M_{0c}^{0.56}
\end{equation}

In our experiments, $M_{0p}$ contributes on average to about 4 $\%$, 6 $\%$ and 2 $\%$ of $M_{0c}$ at Pc = 30, 45 MPa and 60 MPa respectively. This is slightly less that what was found by \citet{passelegue2016dynamic} and \citet{acosta2019precursory} but is typically of the same order of magnitude. Experimental observations may also simply indicates a linear relation between $M_{0p}$ and $M_{0c}$ as given by the slope of 1. 

\subsection{Comparison with natural earthquakes and implications}

The scaling relationship between moment magnitude and corner frequency $M_{0} \propto f_{c}^{3}$ is verified on the scale of crustal faults, induced seismicity or in the laboratory, i.e. for a wide range of moment magnitudes from -8 to 8. \citep{aki1967scaling,abercrombie1995earthquake,hiramatsu2002scaling,prieto2004earthquake,yamada2007stress,kwiatek2011source,yoshimitsu2014magnitude,selvadurai2019laboratory,blanke2021stress}. In line with previous studies, the estimated AE source parameters also satisfy this scaling relationship (Figure 4d). Therefore, foreshocks recorded during the experiments can truly be considered as micro-earthquakes which is determinant for extrapolating the inferences made in the laboratory to the scale of crustal faults. In a sense, foreshocks recorded during the experiments are more similar to natural earthquakes than mainshocks do, since they consist in self-terminating ruptures. As previously mentionned, we found that larger AEs have larger stress-drops. Although we cannot exclude that this feature is related to insufficiently well calibrated acoustic sensors, this feature might also be physically meaningful. Large foreshocks tend to occur closer to stick-slip instability, when the weakening rate is faster due to accelerating slip, which thus may result in larger stress-drops. Assuming that foreshocks highlight the rupture of locked and critically stressed asperities, these asperities become increasingly seismic as fault slip accelerates. This is consistent with observations at the scale of crustal faults. \citet{bouchon2013long} showed that foreshock sequences were more common for interplate than for intraplate earthquakes due to facilitating slow slip phase at plate boundaries. Similarly, \citet{mcguire2005foreshock} have observed that oceanic transform faults with relatively high-slip rates were producing more foreshock sequences. 

\noindent Extending the scaling relationship between $M_{0a}$ and $M_{0p}$ (Figure 10a) to larger pre-seismic moments would rapidly lead to 100 $\%$ of seismic coupling. Fixing $n$ = 4 and taking the experiment conducted at Pc = 45 MPa as an example, $M_{0a}$ would equal $M_{0p}$ for $M_{0p} \approx 10^{4.5}$ N.m which is equivalent to an amount of pre-slip of about 300 $\mu m$. If we assume a ratio of $M_{0p}/M_{0c}$ of about $5\%$, 300 $\mu m$ of pre-slip gives 6 mm of coseismic displacement which is a typical value for a magnitude 2.5-3 earthquake. \citet{tamaribuchi2018characteristics} analysed foreshock(s)-mainshock-aftershock(s) sequences in the JMA catalog over a 20-year period. The authors found that the magnitude of the largest foreshock within a sequence scales with the magnitude of the mainshock but numerous mainshocks are not preceded by foreshocks (at least not by foreshocks of $Mw > 1.0$, the completeness magnitude of the catalog) and, when they do, it is common that the largest foreshock is at least 2 orders of moment magnitude less than that of the mainshock. Therefore, the extrapolation of the scaling relationship we find between $M_{0a}$ and $M_{0p}$ to a larger scale is likely to be invalid. Indeed, in addition to be dictated by the way mainshocks initiate in our experiments, the relationship between $M_{0a}$ and $M_{0p}$ is likely to be controlled by the actual experimental conditions: a constant fault surface, a rapid loading which prevents healing and a smooth planar fault with a low degree of structural complexity (damage zone, fault branches, lithology contrast to cite only a few). Nevertheless, our observations suggest that valuable insights on earthquake nucleation mode, in a specific geological context, can be obtained by examining the relationship between $M_{0a}$ and $M_{0p}$. 

Although the nucleation phase is difficult to image using geodetic measurements, recent observations on well instrumented earthquakes constitute exceptions. The pre-seismic moment was estimated using geodetic for the 2011 $Mw$ 9.0 Tohoku-Oki earthquake \citep{kato2012propagation}, the 2012 $Mw$ 7.6 Nicoya earthquake \citep{voss2018slow}, the 2014 $Mw$ 8.2 Iquique earthquake \citep{socquet20178} and the 2015 $Mw$ 8.4 Illapel earthquake \citep{huang2018slow}. For these earthquakes, $M_{0p}/M_{0c}$ ranges from about 0.4 $\%$ to 3 $\%$ which is very close to our estimates (4 $\%$, 6 $\%$ and 2 $\%$ of at Pc = 30, 45 and 60 MPa respectively). Different forms of (21) were proposed, for instance the one proposed by \citet{acosta2019precursory}, as previously mentioned, or the well-know one $M_{0p} \propto M_{0c}^{0.78}$ proposed by \citet{abercrombie2005can} within the framework of slip-weakening theory and on the basis of seismological observations. Our data alone do not allow us to state on the scaling exponent of (21) but by comparing data from experimental and natural earthquakes, fracture energy must be proportional to co-seismic displacement (i.e, $M_{0p} \propto M_{0c}$) for $M_{0p}/M_{0c}$ to be of the same order over such a large range of moment magnitudes. It is likely that the improvements currently made in seismic/geodetic instrumentation and data processing techniques will make possible to estimate $M_{0p}/M_{0c}$ for a large range of moment magnitudes and thereby will bring new insights to (21).

Comparing the total foreshock moment release $M_{0a}$ with the co-seismic moment release $M_{0p}$ in our experiments, there is up to 7 orders of magnitude difference between $M_{0a}$ and $M_{0c}$, or equivalently 4 orders of magnitude difference in terms of moment magnitude $M_{w}$. At the crustal scale, a case study is the one of the 1999 $M_{w}$ 7.6 Izmit earthquake. While the nucleation of the 1999 Izmit earthquake is still debated (i.e., cascade model \citet{ellsworth2018nucleation}, or preslip model \citet{bouchon2011extended,bouchon2021nucleation}), \citet{bouchon2011extended} inferred the precursory moment $M_{w}$ 7.6 Izmit earthquake from a sequence of repeaters which occurs within the last hour prior to the mainshock. The authors argued that the occurrence of repeaters required a fast reloading of stress and, thereby, manifested the expansion of a nucleation patch. \citet{bouchon2011extended} estimated the pre-seismic moment to be 6 orders of magnitude smaller than the co-seismic moment, a ration abnormally low. However, our data show that seismic coupling may be very low during nucleation and, consequently, that a strict equality between seismic moment released by repeaters and pre-seismic moment is questionable. The seismic moment released by repeaters should be interpreted as the smallest possible value of $M_{0p}$. Moreover, our observations questions the parallel commonly drawn between a lack of detectable seismicity prior to a mainshock and a cascading process. Nucleation process could be too silent in some cases for the nucleation phase to be detected by seismic instruments. Indeed, similar experiments, performed under fluid saturated conditions, led to the absence of detectable foreshock sequences \citep{acosta2019can}, although mainshocks were preceded by a long slow-slip transients. 

Finally, foreshock migration towards mainshock hypocenter is generally attributed to slow-slip propagation \citep{kato2012propagation,ruiz2014intense,kato2014multiple,kato2016foreshock}, stress-transfer (static or dynamic, \citet{ellsworth2018nucleation,yao2020detailed}) or fluid diffusion \citep{moreno20152014,socquet20178}. Here, we proposed an alternative explanation, rarely considered yet theoretically predicted, which is that foreshock migration arises from slip localization, promoted by a rapid weakening rate, onto a fault patch. In practise, foreshock migration due to slip localization would be easily distinguishable from foreshock migration due to fluid diffusion or to slow-slip propagation since fluid-diffusion driven foreshocks migrate as the square root of time, and slow-slip driven foreshocks migrate as slip lines. The question whether slow-slip transients prior to large earthquakes are part of the nucleation process is still debated. As an example, the 2011 $Mw$ $9.0$ Tohoku-oki earthquakes was preceded by slow-slip events but the latter did not propagate with slip (and foreshock rate) acceleration which is kinematically expected in case of a nucleation process. Foreshock migration due to stress-transfer or slip localization share similar spatio-temporal characteristics. To distinguish one from another would require strong constraints on foreshocks size, magnitude and location, provided that foreshock sequence does not stem from the feedback between the two processes as evidenced by the present study and previous experimental works \citep{mclaskey2014preslip,passelegue2017influence,yamashita2021two} and numerical simulations \citep{cattania2021precursory}.

\section{Conclusions}
In this study, we recorded microseismicity generated during stick-slip experiments and analyzed the spatio-temporal dynamics of precursory foreshocks and slip prior to stick-slip instabilities. Using calibrated acoustic sensors, foreshock source parameters were also determined. Our results evidence that the occurrence of foreshocks was driven by fault slip acceleration during the nucleation phase of the upcoming stick-slip instability. Figure 11 summarizes the dynamics of the nucleation phase, during which:

(i) pre-slip on the entire fault was systematically observed preceding failure. Much in agreement with the pioneering work of \citet{dieterich1992earthquake}, slip and stress heterogeneities result in slip localization onto a patch of the fault, which is reflected by foreshock migration towards the epicenter, and a decrease of the b-value from $\sim1$ to $\sim0.5$.

(ii) the foreshock rate is driven by slip velocity. As slip accelerates, so does the foreshock rate, which increases as an inverse Omori-law. An experimental scaling with the nucleation size was found experimentally for the inverse Omori-law, which suggest that indeed, foreshocks are driven by a nucleation process of given length-scale.

(iii) as the fault accelerates, the nucleation size shrinks because of enhanced slip and velocity weakening. At $t=c$ of the inverse Omori-law, the nucleation size has shrank enough (or strain has localized enough, or the weakening rate is fast enough), that we observe the transition from a frictional, 'Dieterich-like' \citep{dieterich1992earthquake}, instability, to that of a fracture, 'Ohnaka-like' \citep{ohnaka2003constitutive}, process.

 (iv) from the ratio between the seismic and the aseismic components of the nucleation phase, we find that this transition from 'slip to crack' also corresponds to the transition between the nucleation phase being almost fully aseismic, towards a cascading process. The question remains opened however on whether the mainshock is truly triggered by a cascade-like process, i.e. whether the mainshock is a foreshock that degenerates by rupturing a patch large - or weak - enough to propagate over the entire fault plane. Recent seismological observations of the self-similarity between small and large earthquakes \citep{ide2019frequent} suggest it could indeed be the case.
 
 \begin{figure}
    \includegraphics[width=\textwidth]{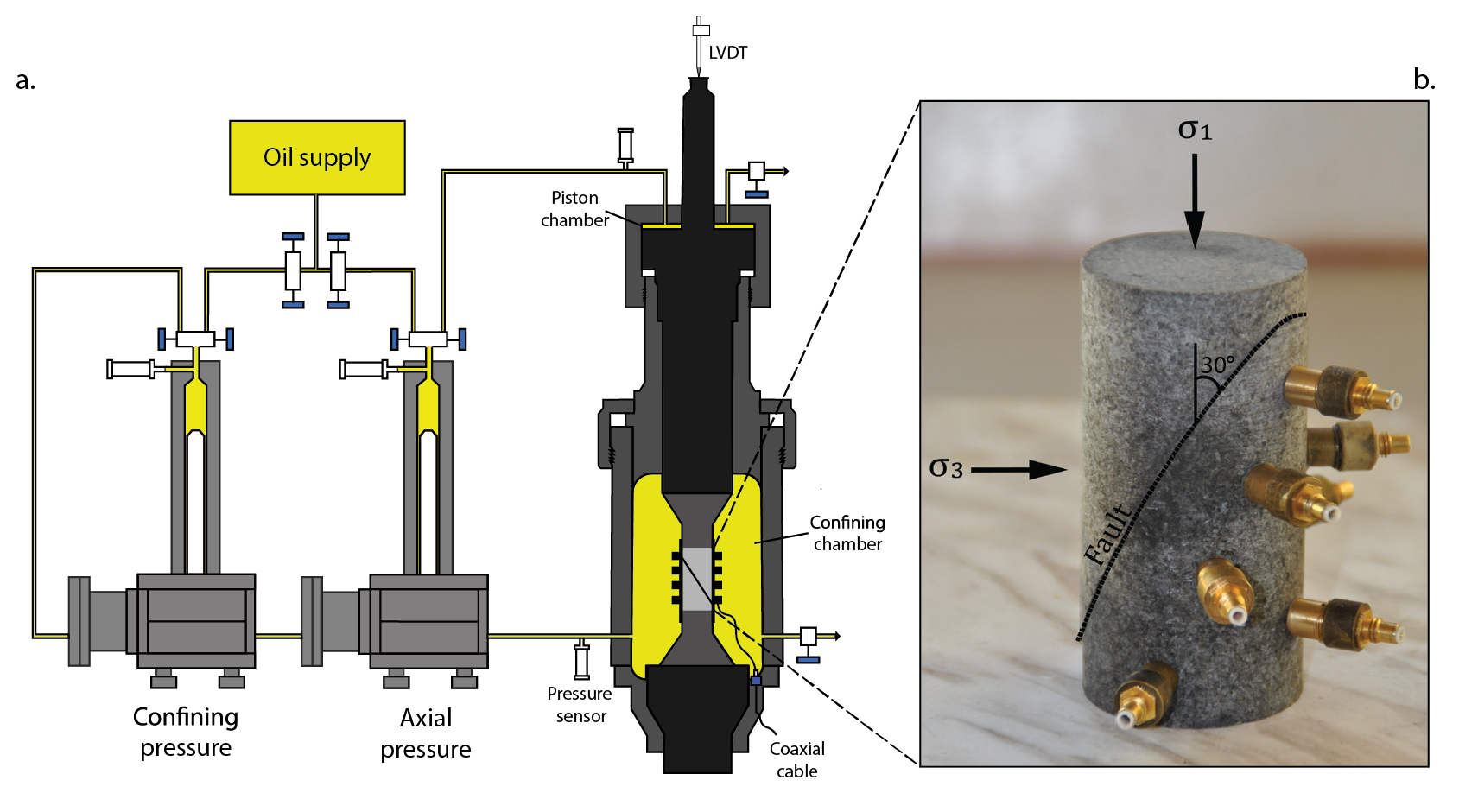}
\caption{\textbf{Triaxial apparatus and rock assemblage.} \textbf{a.} Schematic of the triaxial oil-medium loading cell. Two external servo pumps control axial and radial stresses. Axial displacement is given by the displacement of the piston measured by a LVDT at its top. \textbf{b.} Saw-cut rock specimen used to reproduce laboratory earthquakes. The fault plane is oriented at $30^\circ$ with respect to the principal stress $\sigma_{1}$. Seismic waves generated during the experiments are recorded by acoustic sensors glued at the surface of the sample.}
\label{Figure 1}
\end{figure}

  \begin{figure}
    \includegraphics[width=\textwidth]{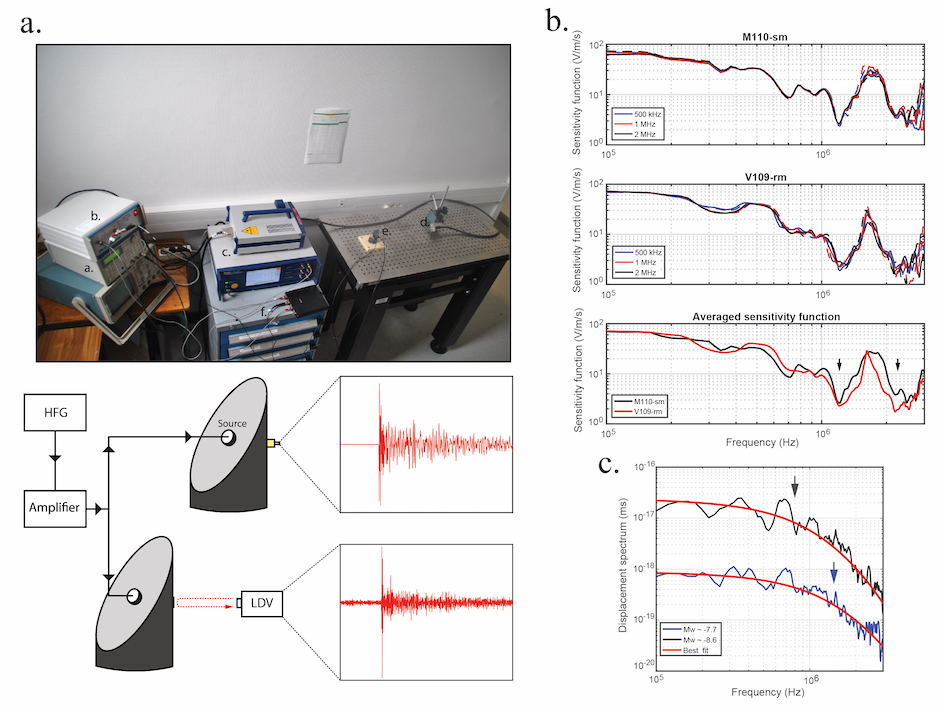}
\caption{\textbf{a.} Photograph and schematic of the experimental set-up used to calibrate the acoustic sensors. \textbf{b.} Calibration results. The two top panels show, for the two transmitter types (M110-sm and V109-rm), the sensitivity functions obtained for the three tested excitation frequencies (500 kHz, 1 MHz and 2 MHz). Transmitters were excited with a 200 V pick-to-pick sinusoidal wave. The sensitivity functions averaged over the excitation frequencies are displayed in the bottom panel. \textbf{c.} Fits of the displacement spectra of a $Mw$ -7.7 AE and a $Mw$ -8.6 AE. Estimated corner frequencies are indicated by the black arrows and are respectively 0.88 MHz and 1.5 MHz.}
\label{Figure 2}
\end{figure}

  \begin{figure}
    \includegraphics[width=\textwidth]{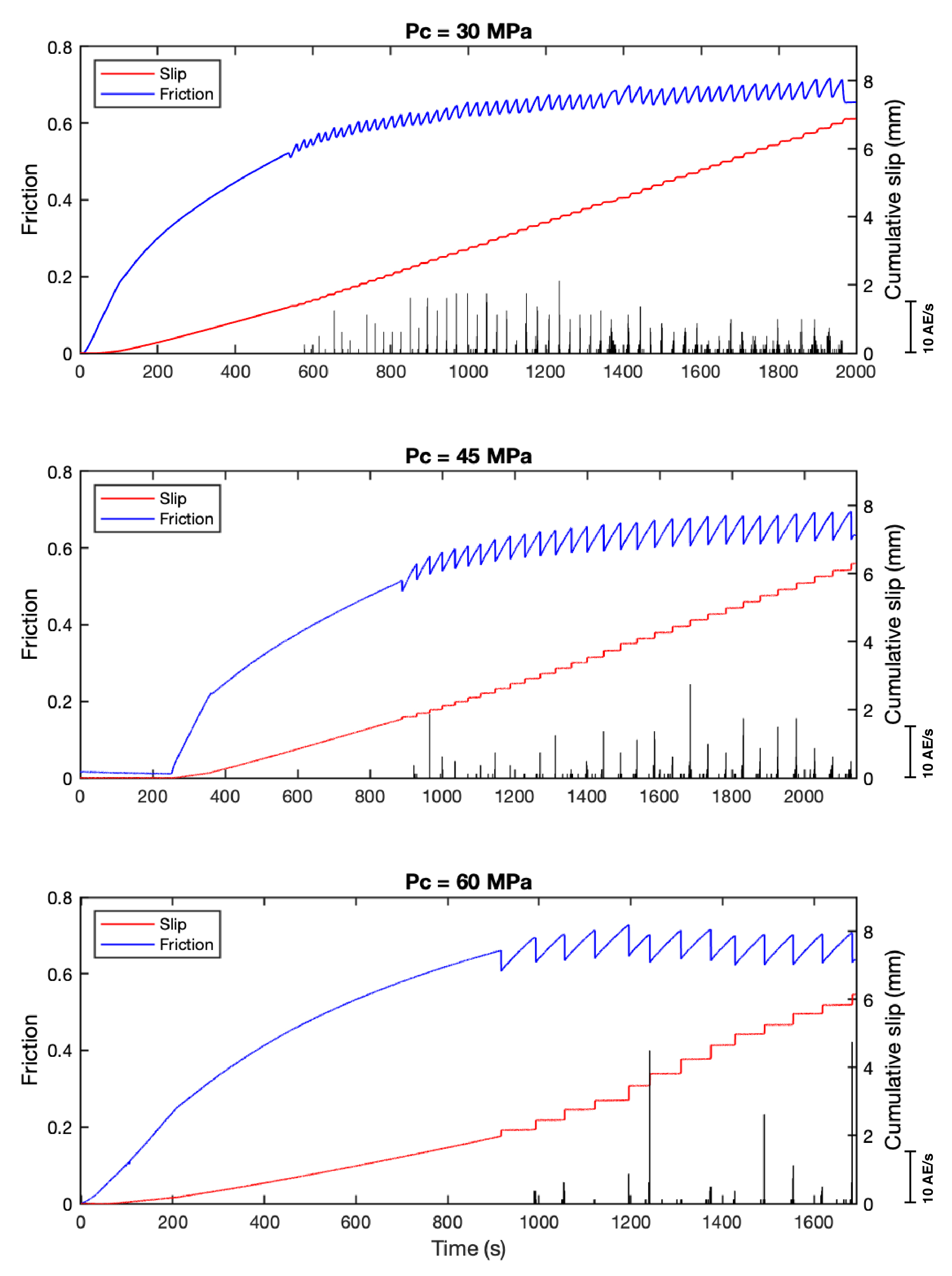}
\caption{Cumulative slip, friction and AE rate during the experiments. AEs were stacked into 1 second bins and cumulative slip was corrected from the elastic deformation of the column apparatus (sample + piston).}
\label{Figure 3}
\end{figure}

  \begin{figure}
    \includegraphics[width=\textwidth]{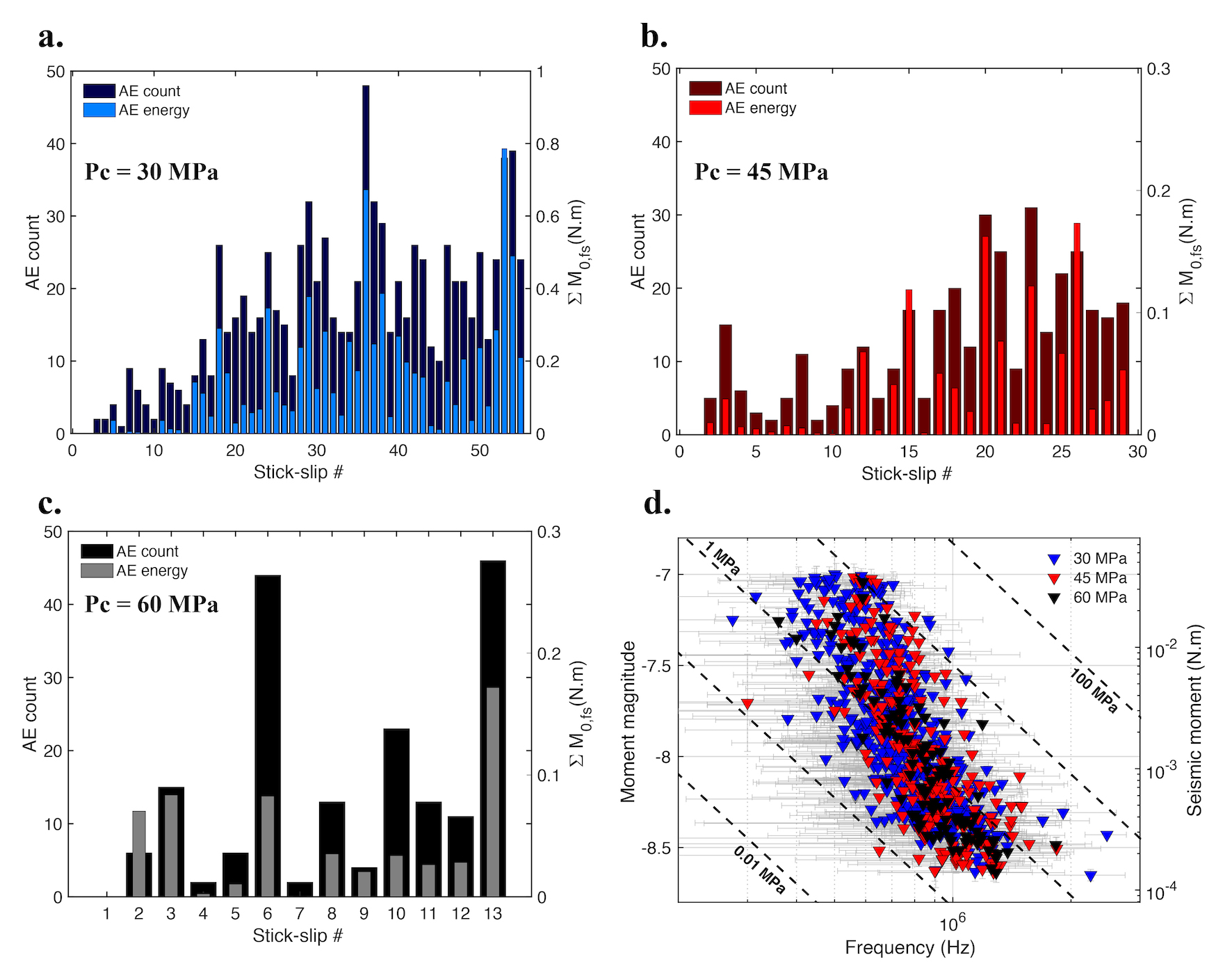}
\caption{\textbf{a-c} Distribution of the number of AEs (dark colors), and the total AE moment release (light colors) per stick-slip cycle for the three experiments. \textbf{d.} Scaling relationship between AE seismic moments $M_{0}$ (or equivalently moment magnitude $M_{w}$) and AE corner frequencies $f_{c}$. Errorbars are shown in light grey. Dashed black lines represent stress drops of 0.01, 0.1, 1, 10, 100 $MPa$ from Madariaga's source model \citep{madariaga1976dynamics}.}
\label{Figure 4}
\end{figure}

  \begin{figure}
    \includegraphics[width=\textwidth]{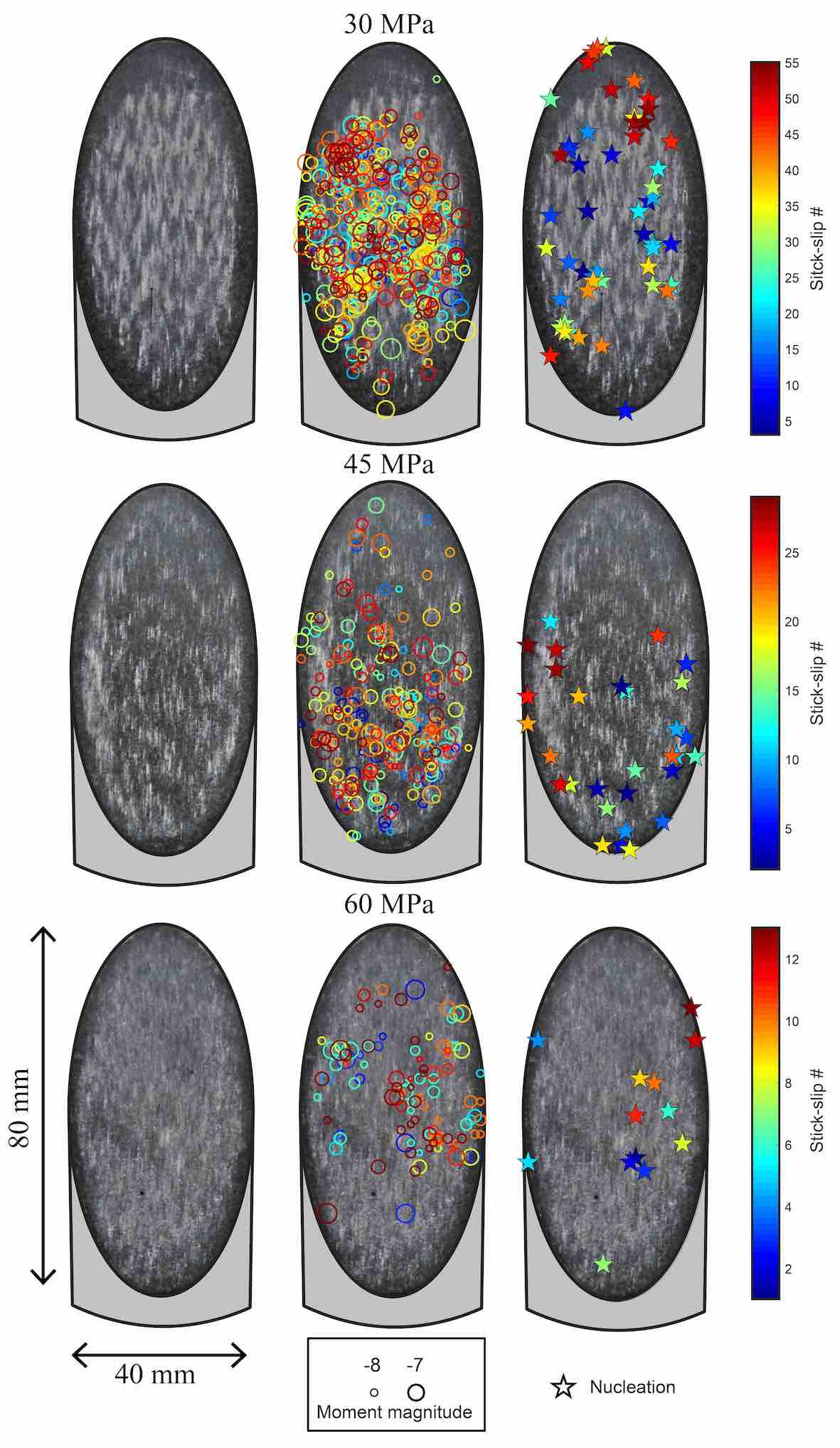}
\caption{Post-experiment fault surfaces conditions (left), AE (center) and SSE hypocenter (right) locations. The size of the circles matches the estimated AE source sizes (assuming a circular shape) and the colorscale refers to the SSE index. Only AEs whose location errors are less than 2-3 mm are reported here.}
\label{Figure 5}
\end{figure}

  \begin{figure}
    \includegraphics[width=\textwidth]{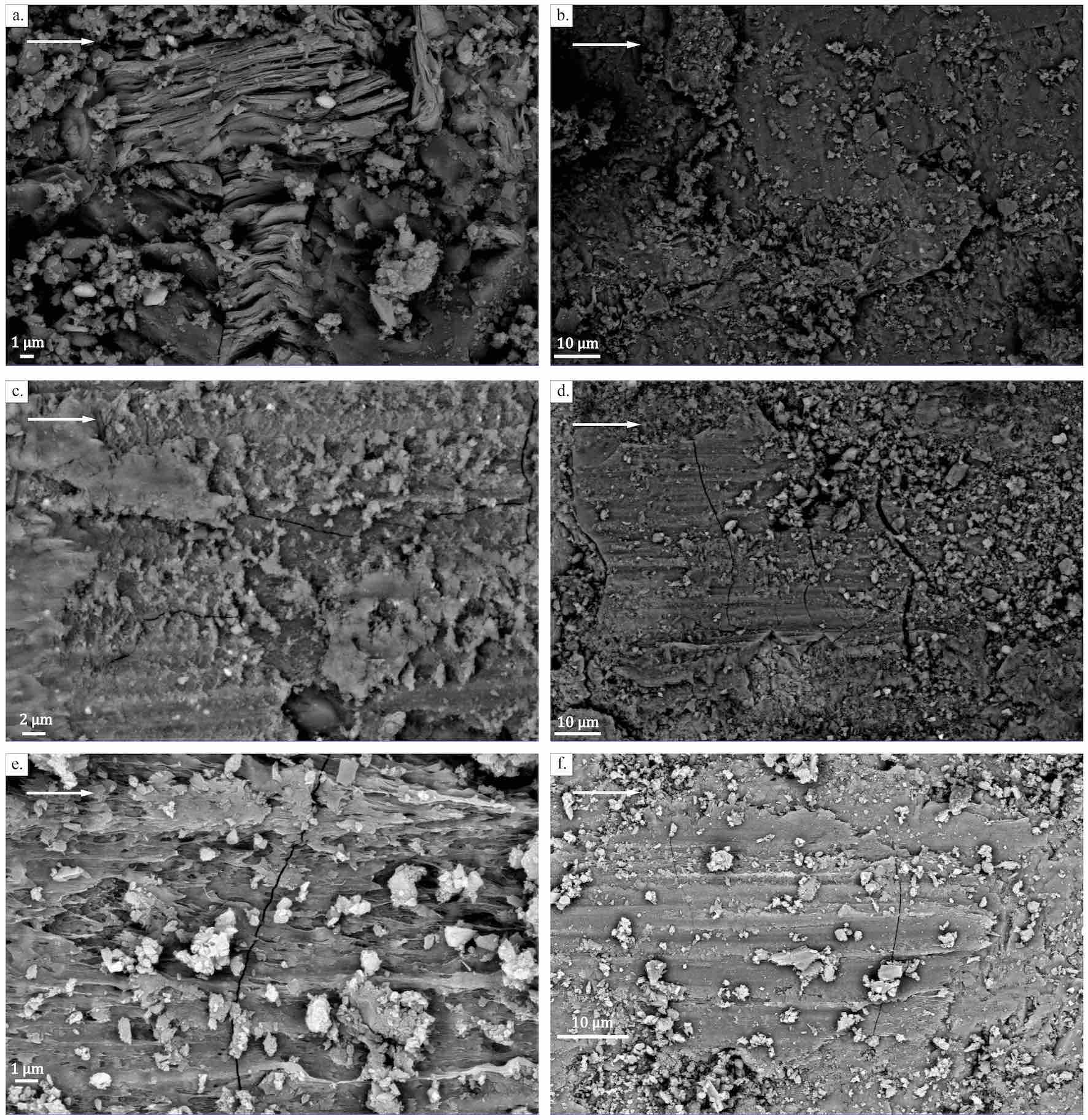}
\caption{Post-experiment microtexture of the fault surfaces under Scanning Electron Microscopy at : \textbf{a-b} Pc = 30 MPa, \textbf{c-d} Pc = 45 MPa and \textbf{e-f} Pc = 60 MPa. White arrows show the direction of sliding. \textbf{a.} Small scale view of gouge particles with various sizes ranging from few $\mu m$ to 100 $nm$. \textbf{b.} Large scale view of \textbf{a.} showing 
gouge patches heterogeneously distributed on a damaged asperity slightly deformed into the direction of sliding. \textbf{c.} Small scale view of amorphous fine gouge particles layer. \textbf{d.} Large scale view of \textbf{c.} showing clusters of smashed gouge particles with sizes up to 10 $ \mu s$. The fault surface presents striations along the sliding direction which suggests plastic deformation during stick-slip events. \textbf{e.} Small scale view of the fault surface showing evidence of partial melting during sliding. A fraction of the gouge particles is trapped into the melt. \textbf{f.} Large scale view of \textbf{e.} showing stretched and elongated surfaces formed due to partial melting and covered with (more) homogeneously distributed gouge particles.}
\label{Figure 6}
\end{figure}

 \begin{figure}
    \includegraphics[width=\textwidth]{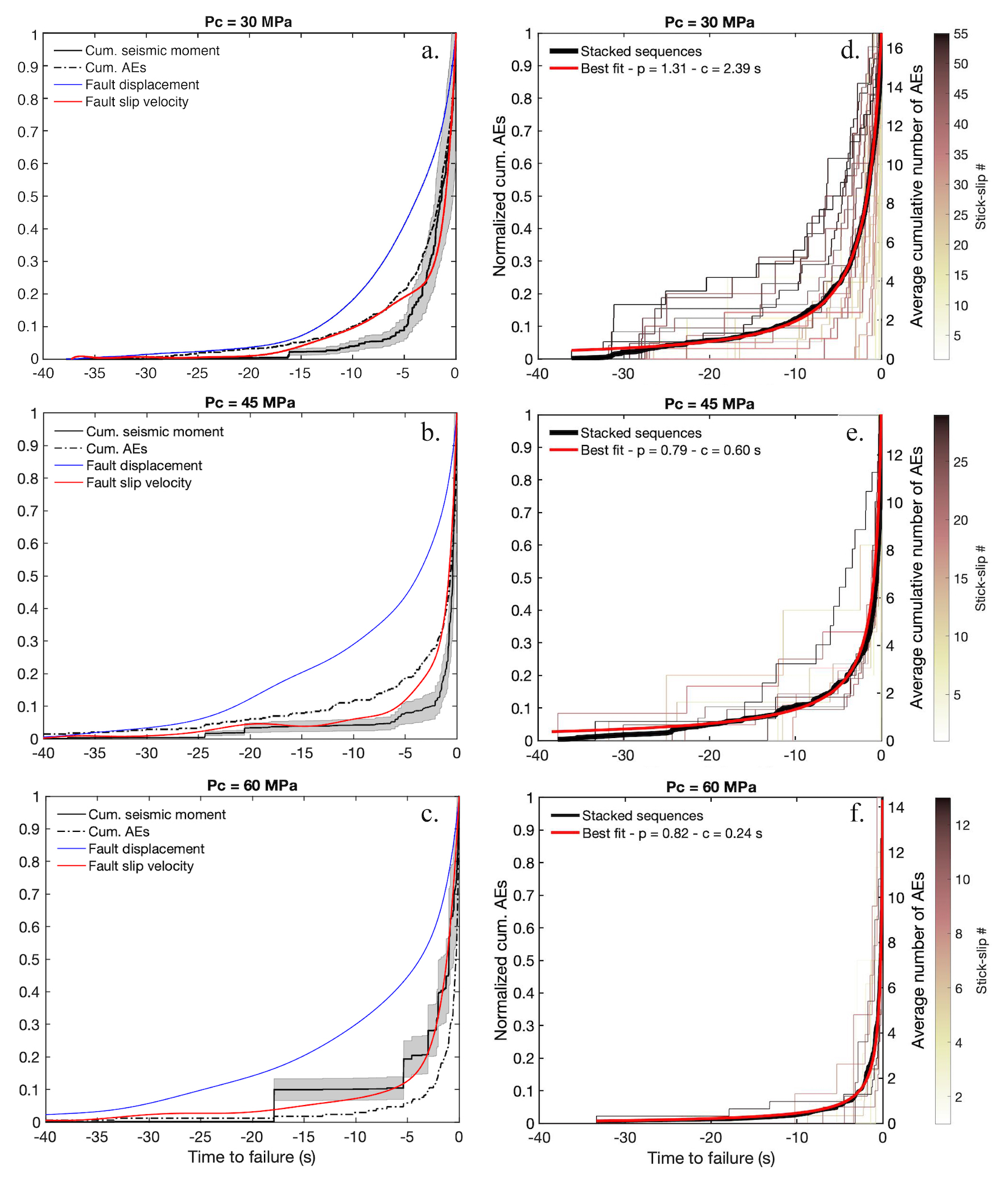}
\caption{\textbf{a-c.} Normalized fault displacement, fault velocity, cumulative number of AEs and cumulative AE moment release in the last 40 s prior to failure at respectively Pc = 30, 45 and 60 MPa. Each curve represents stacking of all SSE sequences. \textbf{d-f.} Inverse Omori fits (red curves) of the stacked cumulative number of AEs (black curves) in the last 40 s prior to failure at respectively Pc = 30, 45 and 60 MPa. The color curves display the individual precursory AE sequences, with the colorscale referring to the SSE index.}
\label{Figure 7}
\end{figure}

 \begin{figure}
    \includegraphics[width=\textwidth]{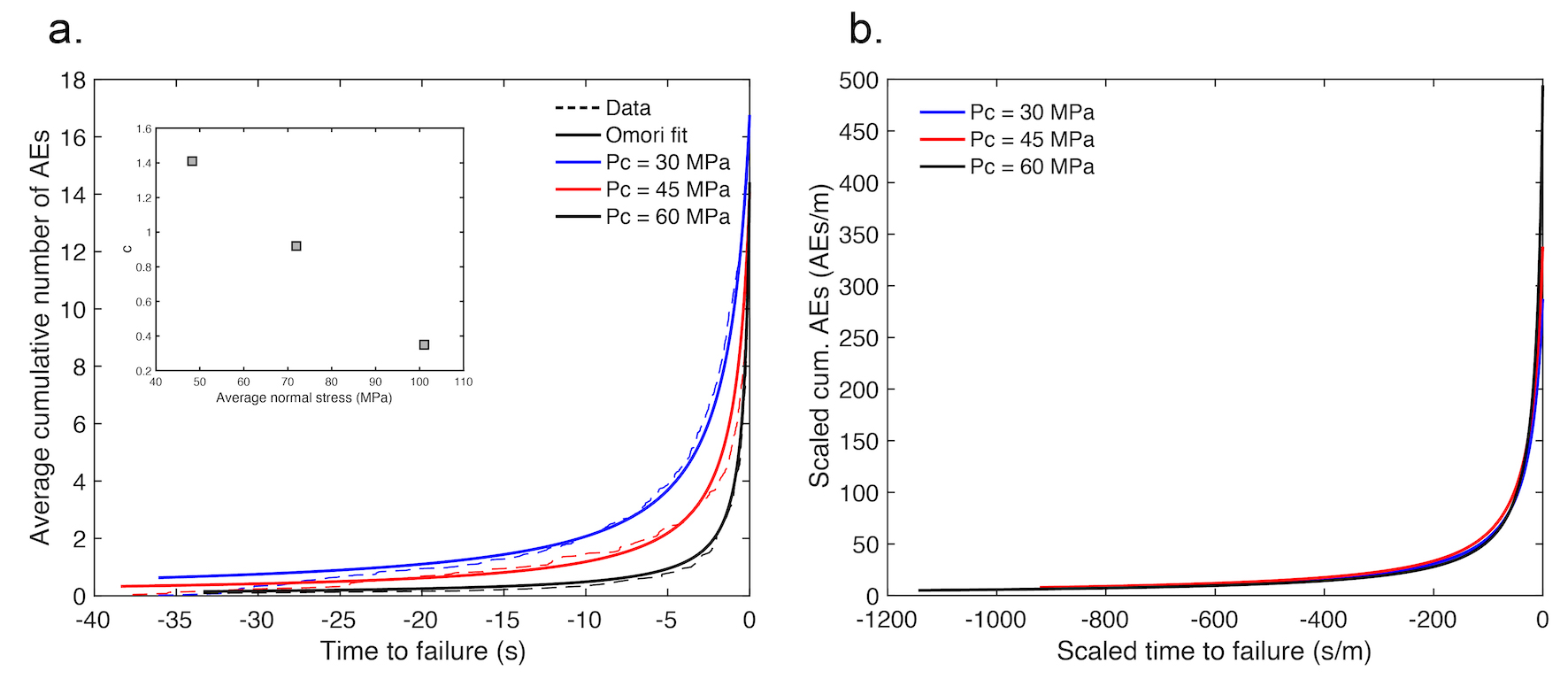}
\caption{Scaling of the inverse Omori law. \textbf{a.} Best inverse Omori fits (solid lines) of the stacked cumulative number of AEs (dashed lines) obtained by imposing $p$= $1$ at respectively Pc = 30, 45 and 60 MPa. The inset shows the linear relationship between $c$ and normal stress with $c$ = $1.41$, $0.92$ and $0.35$s at respectively Pc = 30, 45 and 60 MPa. \textbf{b.} Inverse Omori fits scaled with the nucleation length $L_{b}$ at respectively Pc = 30, 45 and 60 MPa. The curves collapse by normalizing the productivity $K$ (y-axis) and the time to failure (x-axis) by $L_{b}$.}
\label{Figure 8}
\end{figure}

 \begin{figure}
    \includegraphics[width=\textwidth]{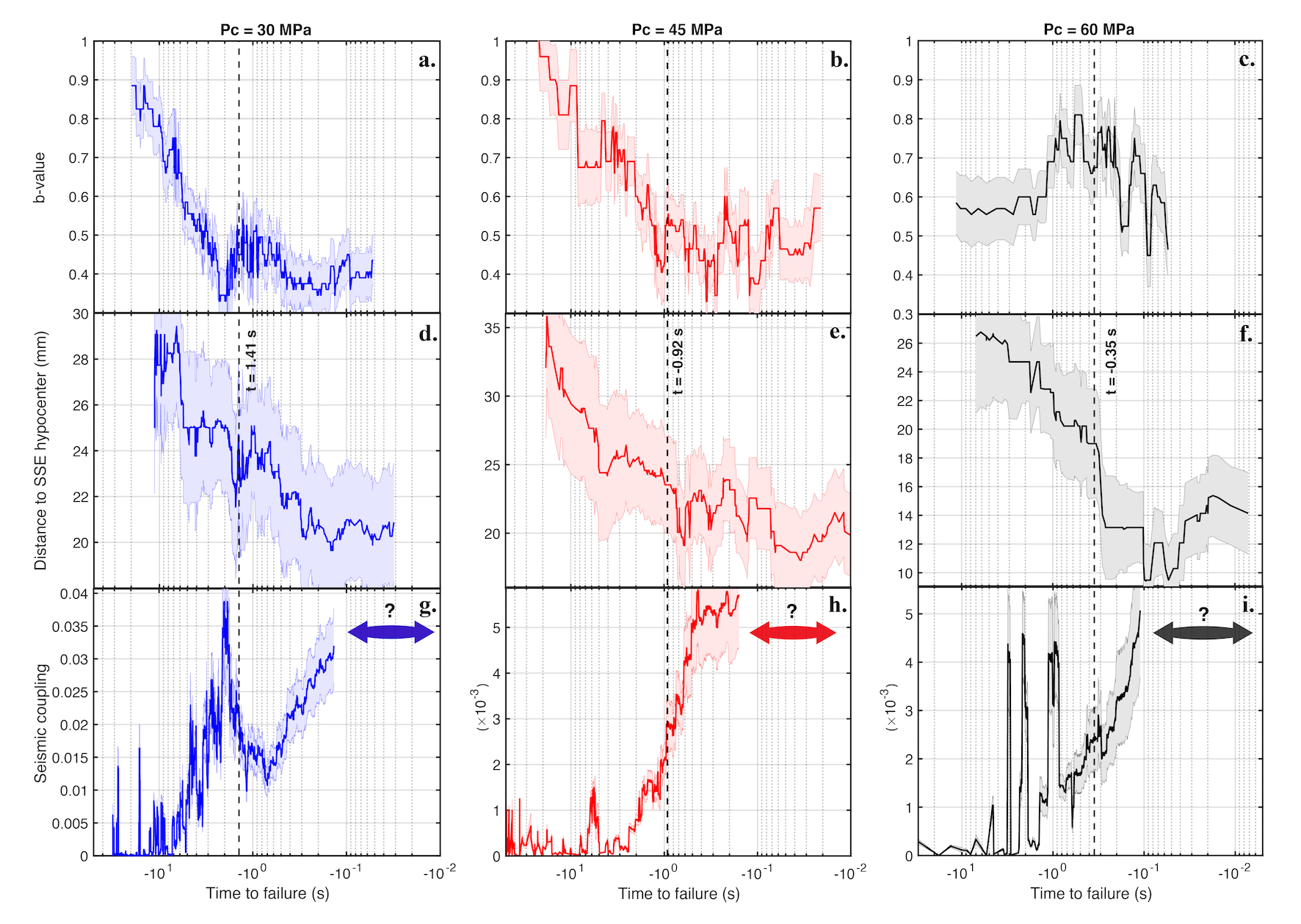}
\caption{b-value, precursory AE migration and seismic coupling as a function of the logarithm of time to failure. Dashed lines indicate the parameter $c$ of the best Omori inverse models obtained by imposing $p$ = $1$. \textbf{a-c.} b-value estimated within a moving window containing 100 AEs at Pc = 30, 45 MPa and 50 AEs  at Pc = 60 MPa. Shaded areas correspond to the 90$\%$ confidence intervals. \textbf{d-f.} Average distance to SSE hypocenter of the precursory AEs computed within a moving window containing 50 AEs at Pc = 30 and 45 MPa and 25 AEs  at Pc = 60 MPa. Shaded areas correspond to the $\pm$ one standard deviation intervals. \textbf{g-i.} Seismic coupling computed within a moving window of duration equal to 1$\%$ of the duration of the stacked precursory AE sequence. Shaded areas display the range of uncertainty computed from foreshock magnitude uncertainties. Double headed arrows indicate the time range for which seismic coupling cannot be estimated due to a temporal resolution of $\sim$ 0.1s.}
\label{Figure 9}
\end{figure}

 \begin{figure}
    \includegraphics[width=\textwidth]{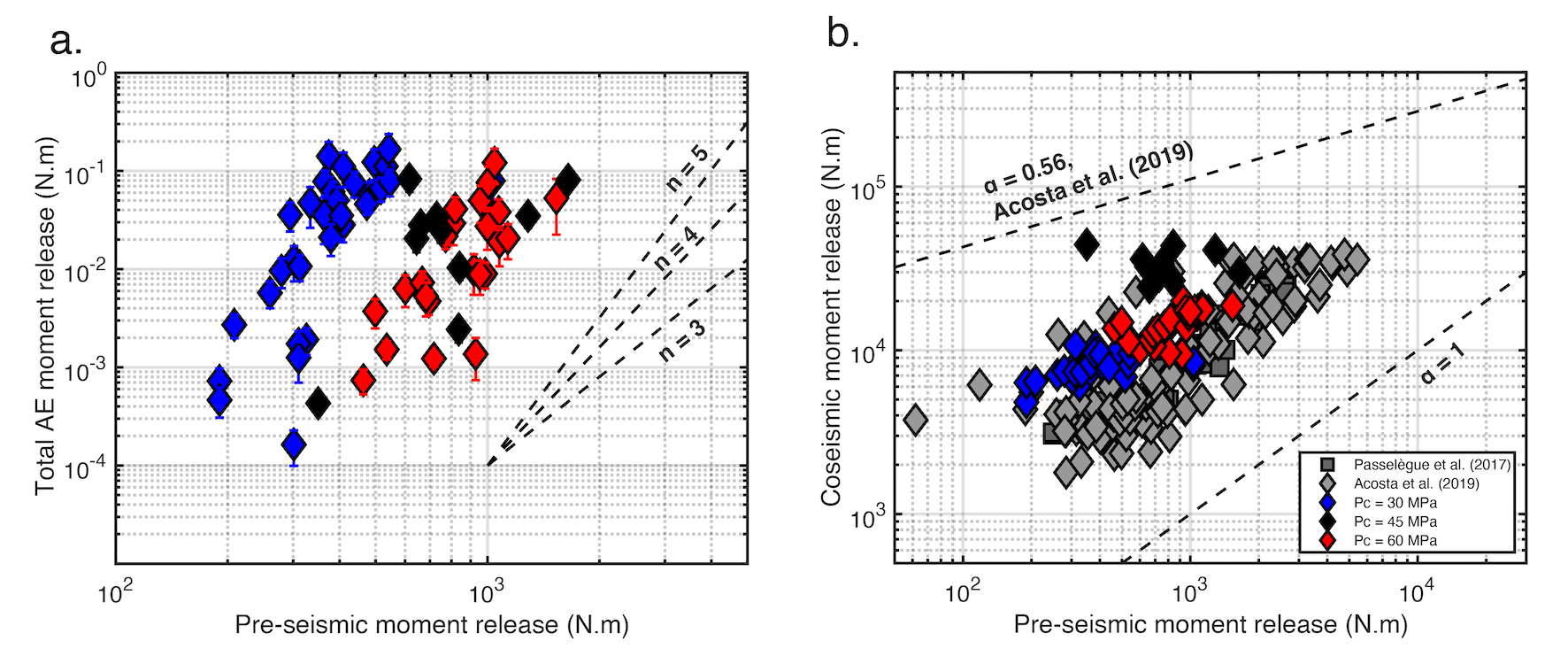}
\caption{\textbf{a.} Total AE moment release as a function of pre-seismic moment release. Each marker represents one SSE. The black dashed lines show the slopes for power law type relations of exponent $n$ = 4, 5 and 6. \textbf{b.} Co-seismic moment release as a function of pre-seismic moment release. As a comparison, the data from the present study (diamond symbols) are plotted together with the data (grey squares and circles) from two other experimental studies \citep{passelegue2017influence,acosta2019precursory}. The black dashed line with slope 0.56 corresponds to the scaling law between pre-seismic moment release $M_{p}$ and co-seismic moment release $M_{c}$ proposed by \citet{acosta2019precursory}. A linear relation between both quantity is given by the black dashed line with slope 1.}
\label{Figure 10}
\end{figure}

\begin{figure}
    \includegraphics[width=\textwidth]{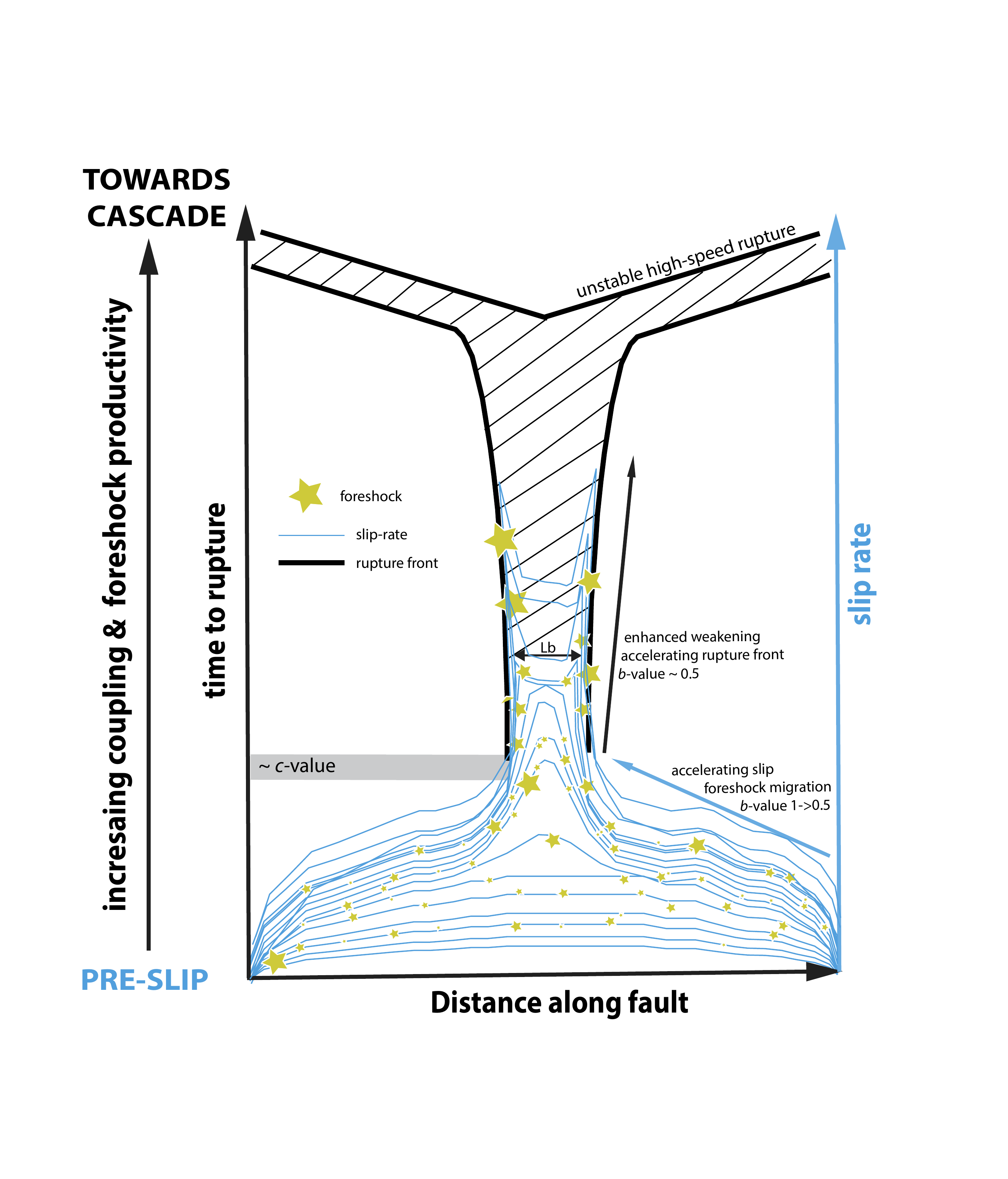}
\caption{\textbf{Schematics of nucleation phase dynamics.} Slip and stress heterogeneities result in slip localization onto a patch of the fault, which is reflected by foreshock migration towards the epicenter. At t = c of the inverse Omori-law, we observe the transition from a frictional, 'Dieterich-like' \citep{dieterich1992earthquake}, instability, to that of a fracture, 'Ohnaka-like' \citep{ohnaka2003constitutive}, process.}
\label{Figure 11}
\end{figure}

\clearpage

%
%
%
%
%
%
%
%

\acknowledgments
This work was funded by the European Research Council grant REALISM (2016‐grant 681346). H. S. B. acknowledges the European Research Council grant PERSISMO (grant 865411) for partial support of this work. The authors declare that they have no competing financial interests. All data are available online: \url{https://doi.org/10.4121/14619222.v1}.

\clearpage
\section*{Supporting Information for "Dominantly Aseismic Nucleation of Laboratory Earthquakes: A Quantitative Investigation"}

\noindent\textbf{Introduction}

We provide to the reader additional details concerning (i) the calculation of along fault displacement, shear stress and normal stress and (ii) the methodology that was adopted for acoustic sensors calibration and, acoustic emission detection and localization.

\noindent\textbf{Text S1.}

\section*{Calculation of along fault displacement, shear stress and normal stress}
The displacement measured at the top of the axial piston by the LDV includes the shortening of the rock sample and the axial column. Both the rock sample and the axial column will deform elastically during loading. The displacement along the fault surface is obtained by correcting the axial deformation of the elastic deformation of the sample and the axial piston such as :

\begin{equation}
    \epsilon_{ax}^{FS}=\epsilon_{ax}^{WS}-\frac{\Delta\sigma}{E_{ap}}-\frac{\Delta\sigma}{E_{s}}
\end{equation}
 
 where $\epsilon_{ax}^{FS}$ is the axial deformation of the fault surface, $\epsilon_{ax}^{WS}$ is the axial deformation of the whole system measured at the top of the axial piston, $\Delta\sigma$ is the differential stress, and $E_{ap}$ and $E_{ap}$ are the rigidity of the apparatus and the sample respectively. The displacement along the fault surface is given by the projection on the fault of $\epsilon_{ax}^{FS}$ times the length of the sample. \\
The shear stress $\tau$ and the normal stress $\sigma_{n}$ acting onto the fault can be expressed as:

\begin{equation}
\tau=\frac{\sigma_{1}-\sigma_{3}}{2}.sin(2\theta) 
\end{equation}
and
\begin{equation}
\sigma_{n}=\frac{\sigma_{1}+\sigma_{3}}{2}+\frac{\sigma_{1}-\sigma_{3}}{2}.cos(2\theta) 
\end{equation}
where $\sigma_{1}$ and $\sigma_{3}$ correspond to the axial stress and the confining pressure respectively and $\theta$ to the angle between the direction of $\sigma_{1}$ and the fault plane ($\theta = 30^{\circ}$).

\noindent\textbf{Text S2.}

\section*{Acoustic sensors calibration}

After several trials, acoustic sensors calibration was performed on one of the half cylinders of a saw-cut sample of Indian metagabbro used during the experiments to be as consistent as possible to experimental conditions. The sample was first fixed to an optical table to filter parasite signals. A broadband transducer was affixed to the center of the fault interface which had been preliminarily rectified to ensure good contact. Using a high frequency generator, an amplified step voltage was applied to the transducer. The vibration of the opposing sample surface was measured at 10 $MHz$ sampling rate by the acoustic sensor in the form of a voltage and relayed to a digital oscilloscope. To make sure that the LDV and the acoustic sensor would sample surface vibrations at the exact same location, the position of the acoustic sensor was pointed by the laser beam. Then the acoustic sensor was removed and the same procedure was repeated with the LDV which was set to measure particle velocity. Figure S1 shows a photograph and a schematic view of the experimental set-up. The acoustic sensor's instrumental response (also called the sensitivity function) was obtained from the ratio between the spectrum of the waveform recorded by the acoustic sensor and the spectrum of the velocity waveform recorded by the LDV. The unity of the sensitivity function is $V/m.s^{-1}$.\\
Figure S2 shows an example of the waveforms recorded by the LDV and the acoustic sensor with their respective spectra for the two types of sources. The black double arrow indicates the 50 $\mu s$ long time window used to estimate the spectra. The length of the time window was chosen according to the one used to estimate acoustic emissions seismological parameters. This is required if we don't want to introduce artificial features when deconvolving acoustic emission spectra of the acoustic sensor's intrumental response. It is interesting to note that the spectra are really different with respect to the source. The source M110-sm shows a spike of energy centered around 1 $MHz$ while the source V109-rm generated more energy at lower frequencies. This is convenient because it offers a path to judge of the robustness of the calibration results relative to the frequency content of the source. Moreover, both types of sources have produced sufficiently high-frequency energy to confirm that the acoustic sensors are sensitive to frequencies up to at least 2.5 MHz.\\

\begin{figure}
\centering
    \includegraphics[width=0.8\textwidth]{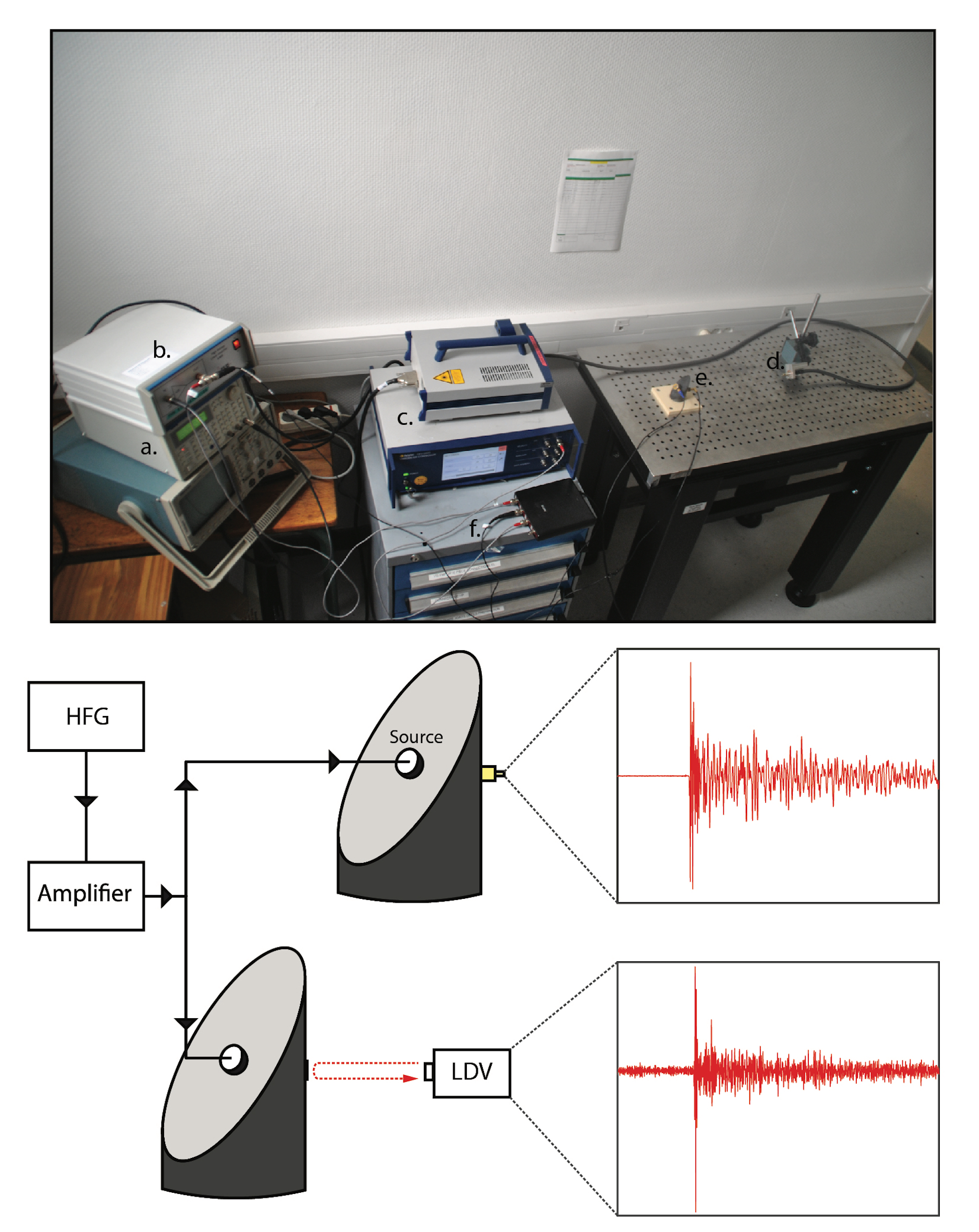}
\caption{\textbf{Top}. Photograph of the experimental set-up used for acoustic sensors calibration. a. High frequency generator (HFG) . b. Amplifier. c. Laser vibrometer acquisition system. d. Laser beam. e. Rock sample with the acoustic sensor and the source glued on. f. Digital oscilloscope. \textbf{Bottom}. Schematic view of the calibration procedure. The source is positioned at the center of the fault and subject to an input voltage. Surface vibrations of the opposing side are recorded by the acoustic sensor first and then by LDV.}
\label{Figure S1}
\end{figure}

\begin{figure}
    \includegraphics[width=\textwidth]{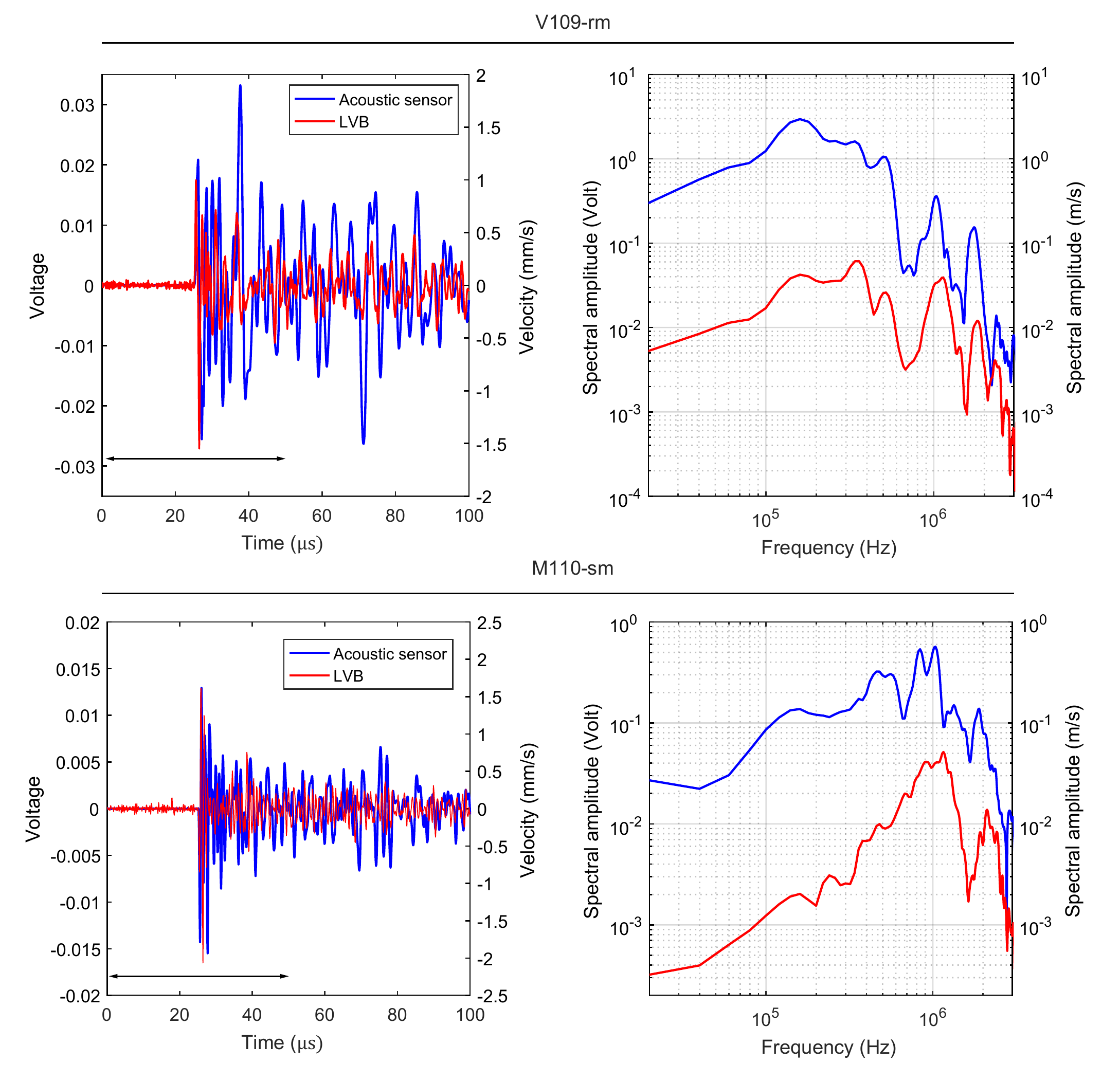}
\caption{Example of voltage and velocity measurements for the two types of sources and the estimated spectra. The time window used to estimate the spectra is indicated by the black double arrow. This time window is 50 $\mu s$ long and is centered to the first P-wave arrival. }
\label{Figure S2}
\end{figure}

\noindent\textbf{Text S3.}
\section*{Acoustic emission detection}

Acoustic emissions generated during the experiments were searched within the continuous waveforms by using a 406.9 $\mu s$ sliding time window with 102.4 $\mu s$ overlap. For a specific time window, acoustic waveforms were saved for all channels if 3 or more channels had recorded an amplitude higher than a particular set threshold (i.e, one specific amplitude threshold for one acoustic sensor). As we wanted to record as much acoustic emissions as possible , the amplitude thresholds were close to the noise level. Therefore, all saved acoustic waveforms were visually inspected to sort between acoustic emissions and noise.

  \begin{figure}
    \includegraphics[width=\textwidth]{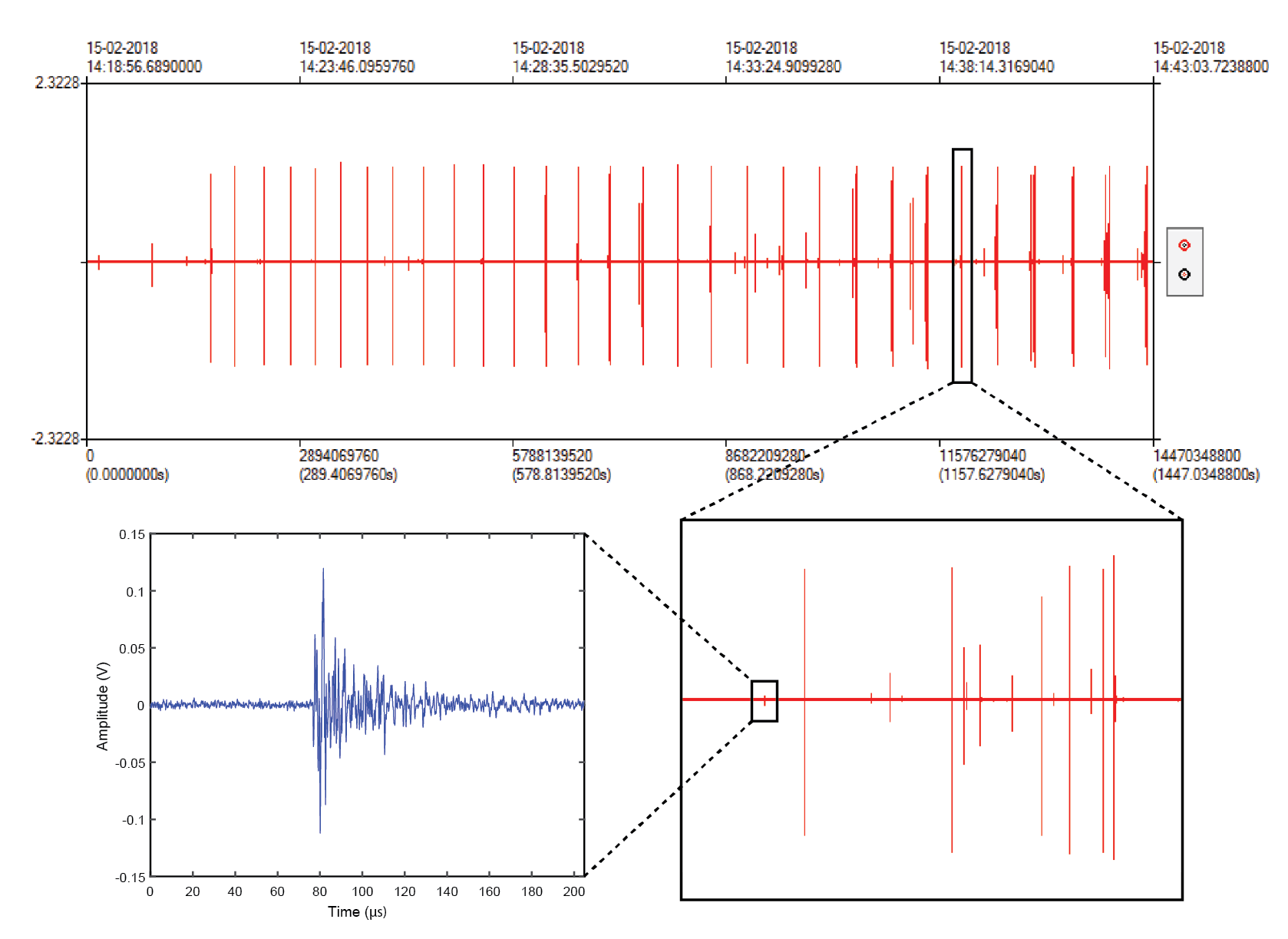}
\caption{\textbf{Continuous recording.} On top is displayed an example of one continuous acoustic recording. The large and regular spikes correspond to the stick-slip events. Close to nucleation (bottom right), the microseismicity is intensifying. Acoustic emissions (bottom left) are searched within the continuous waveform using a sliding time window.}
\label{Figure S3}
\end{figure}

\noindent\textbf{Text S4.}

\section*{Acoustic emission source localization}

Acoustic emissions and stick-slip nucleation were localized according to first P-wave arrivals with $0.1 \mu s$ resolution. Possible positions were restricted to the fault plane. Let ($X_{i},Y_{i},Z_{i}$) indicate the  spatial coordinates of a position onto the fault plane and ($X_{s},Y_{s},Z_{s}$) the spatial coordinates of the acoustic sensors. We used a "double difference algorithm" to locate acoustic emissions or stick-slip nucleation. Acoustic emissions and nucleation positions were obtained by minimizing the L2 norm (least-square) of the sum of the differences between observed and theoretical arrival time differences of all possible pairs of acoustic sensors. This is mathematically expressed as : 

\begin{equation}
    \Delta t(C_{p},i)=2\times\frac{\sum_{cref=1}^{n}\sum_{c\ne cref}^{n-1}\sqrt{(\Delta t_{cref,c,i}^{obs} - \Delta t_{cref,c,i}^{t})^{2}}}{n(n+1)}
\end{equation}

where $C_{p}$  is the P-wave velocity and with:
\begin{equation}
    \Delta t_{cref,c,i}^{t} = (t_{i,cref}-t_{i,c})
\end{equation}
Theoretical travel times $t_{i,c}$ are calculated by the expression : 

\begin{equation}
   t_{i,c}= \frac{\sqrt{(X_{i}-X_{c})^{2}+(Y_{i}-Y_{c})^{2}+(Z_{i}-Z_{c})^{2}}}{C_{p}}
\end{equation}

\end{document}